\documentclass[a4paper,twocolumn,showkeys,floatfix,aps,prd,amsmath,amssymb,nofootinbib,preprintnumbers,superscriptaddress]{revtex4}

\usepackage{times,amsfonts}
\usepackage{natbib}
\usepackage[english]{babel}
\usepackage[T1]{fontenc}
\usepackage[latin2]{inputenc}
\usepackage{graphicx}
\usepackage{color}
\usepackage{tabularx}
\usepackage{mathrsfs}

     \usepackage{color}  \definecolor{myred}{rgb}{0.7,0.0,0.2}

\bibpunct{(}{)}{;}{a}{}{,}
\usepackage{aas_macros}

\usepackage{hyperref}
\usepackage[hyphenbreaks]{breakurl}
\usepackage{url} 

\makeatother

\newcommand\sampleHS{{\em ``targeted''~}}
\newcommand\sampleDF{{\em ``blind''~}}
\newcommand\haloPlotWidth{0.20}

\renewcommand{\d}[1]{\ensuremath{\operatorname{d}\!{#1}}}

\providecommand\gtapprox{\,\lower.6ex\hbox{$\buildrel >\over \sim$} \, }
\providecommand\ltapprox{\lesssim}
\providecommand\ltapprox{\,\lower.6ex\hbox{$\buildrel <\over \sim$} \, }

\newcommand{\Beginruledtabular}{\begin{ruledtabular}}
\newcommand{\Endruledtabular}{\end{ruledtabular}}

\newcommand{\BeginruledtabularSec}{}
\newcommand{\EndruledtabularSec}{}

\begin{document}

\title{The comptonization parameter from simulations of
single-frequency, single-dish, dual-beam, cm-wave observations
of galaxy clusters and mitigating CMB confusion using the
Planck sky survey}

\author{Bartosz Lew} \email[]{blew@astro.uni.torun.pl }
\author{Boudewijn F. Roukema}
\affiliation{Toru\'n Centre for Astronomy,  \protect{Faculty of Physics, Astronomy and Informatics, Grudziadzka 5,}
Nicolaus Copernicus University, ul. Gagarina 11, 87-100 Toru\'n, Poland}

\date{Oct 10, 2016}

\begin{abstract}
Systematic effects in dual-beam, differential, radio
observations of extended objects are discussed in the context of
the One Centimeter Receiver Array (OCRA).  We use simulated
samples of Sunyaev--Zel'dovich (SZ) galaxy clusters at low
($z<0.4$) and intermediate ($0.4<z<1.0$) redshifts to study the
implications of operating at a single frequency (30~GHz) on the
accuracy of extracting SZ flux densities and of reconstructing
comptonization parameters with OCRA. We analyze dependences on
cluster mass, redshift, observation strategy, and telescope
pointing accuracy.  Using {\em Planck} data to make primary
cosmic microwave background (CMB) templates, we test the
feasibility of mitigating CMB confusion effects in observations
of SZ profiles at angular scales larger than the separation of
the receiver beams.

\end{abstract}

\keywords{
Sunyaev-Zeldovich effect --
cosmological simulations --
galaxy clusters --
radio surveys --
methods: observational
}

\maketitle

\section{Introduction}
\label{sec:intro}
Over the past several years many dedicated experiments
have been used to detect
the Sunyaev--Zel'dovich (SZ) effect \citep{Sunyaev1970}
from galaxy clusters at radio wavelengths [e.g.,
Berkeley-Illinois-Maryland Association (BIMA) \citep{Dawson2006};
Combined Array for Research in Millimeter-wave Astronomy (CARMA) \citep{Muchovej2012, Mantz2014};
the South Pole Telescope (SPT) \citep{Reichardt2013};
the N\'eel IRAM KIDs Array (NIKA) \citep{Adam2014};
the Atacama Pathfinder EXperiment Sunyaev--Zel'dovich Instrument (APEX-SZ) \citep{Dobbs2006, Bender2016};
the Arcminute Microkelvin Imager (AMI) \citep{Zwart2008, Rumsey2016};
{\em Planck} Surveyor  \citep{PlanckCollaboration2011};
the Atacama Cosmology Telescope (ACT) \citep{Hasselfield2013};
Array for Microwave Background Anisotropy (AMiBA) \citep{Lin2016}].
Within the next few years, new observational facilities
will become operational and will search for galaxy clusters,
complementing the galaxy cluster census across
the Universe
[e.g.,
New IRAM KID Array 2 (NIKA2) on
the Institut de Radio Astronomie Millimetrique 30~m telescope \citep{Calvo2016}].

The One Centimeter Receiver Array (OCRA)
\citep{Browne2000,Peel2011} is one of the
experiments capable of detecting the SZ effect at 30~GHz using
beam-switching radiometers installed on a 32-meter radio
telescope \citep{Lancaster2007,Lancaster2011}.
OCRA will be mostly sensitive to SZ clusters with virial size $> 3'$
and hence to clusters at redshifts in the range $0.1 < z < 0.5$ and with
masses $M_{\rm{vir}} > 3\times 10^{14} M_{\odot}/h$ \citep{Lew2015}.
However, a single
frequency, beam-switching system may suffer from confusion with
the primordial cosmic microwave background (CMB) or suffer from
systematic error when observing extended sources.

Confusion effects due to the CMB were investigated in detail by
\cite{Melin2006} for AMI, SPT and {\em Planck} Surveyor.  It was
found that for single frequency instruments, such as AMI (a
15~GHz interferometer), the photometric accuracy that contributes
to the accuracy of the reconstructed comptonization parameter is
strongly limited due to primary CMB confusion.

In \cite{Lew2015} the impact of CMB flux density confusion at 30
GHz was investigated, in particular for the OCRA/RT32 (32~m Radio
Telescope in Toru\'n, Poland) experiment.  It was found that the
$1\sigma$ thermal SZ (tSZ) flux density uncertainty due to CMB
confusion should be of the order of $10\%$ for the range of
clusters detectable with OCRA.  However, in that work, the impact
on the reconstructed comptonization parameter in the presence of
the CMB and radio sources was not calculated directly for the
case of dual-beam differential observations.

The $\approx 3'$ separation of OCRA beams is very effective in
CMB removal, but large correcting factors are required to
compensate for the missing SZ signal (after accounting for point
sources) \citep{Lancaster2007}.  Thus, there is a trade-off
between compromising photometry by the primary CMB signal versus
losing flux due to the differential beam pattern.  In between
these extremes, there should exist an optimal separation of
differential beams that would need to be defined by criteria that
aim to maximize CMB removal and minimize SZ flux density removal.

In this paper, we reconsider the issue of systematic effects on
the reconstructed comptonization parameter from single frequency,
beam-switched observations performed with a cm~wavelength
radiometer.  We consider a particular instrumental setting for
the OCRA/RT32 experiment and an extension to the standard
observation scheme that previously involved only the angular
scales defined by the receiver feeds. The extension adds
additional beam pointings that map cluster peripheries, further
from the central core than the initial pointings.

The {\em kinetic} SZ (kSZ) may significantly modify the
brightness of the cluster peripheries that are integrated with
the reference beam. The significance of this effect depends on a
combination of the peculiar velocities of the intra-cluster
medium (ICM) and internal gas clumps, but at cm wavelengths, the
kSZ only weakly modifies the central brightness.

With dual-beam observations, the reference beam background
coverage improves while integrating along arcs around the cluster
center as the field of view (FOV) rotates.  However, due to the
small angular size of the arcs, the chance of zeroing the average
background may be low, depending on the alignment with the CMB
pattern. We investigate the significance of this effect depending
on observational strategy.

For experiments limited by the size of the focal plane array the
integration time required to generate a radio map and to probe
the outer regions of a galaxy cluster is significant and can make
the observation prohibitive.  Therefore, previously, the method
of reconstructing comptonization parameters from OCRA
observations of cluster central regions required inclusion of
X-ray luminosity data in order to find the best fitting
$\beta$-model for each cluster, and correction for the SZ power
lost due to the close beam separation.  However, this approach
relies on the cluster model assumptions and makes the radio SZ
measurements dependent on X-ray measurements of the cluster.
Another possible approach is to observe SZ clusters out to larger
angular distances but retain averaging over a range of
parallactic angles.  This is done at the cost of incurring extra
noise due to weak tSZ in cluster peripheries and stronger
systematic effects due to CMB.

An OCRA-SZ observational program is presently underway. In
support of this and similar efforts, we also investigate the
possibility of mitigating CMB confusion by using the available
{\em Planck} data.  Finally, we calculate the astrometric
pointing and tracking accuracy requirements needed to attain a
given accuracy in flux density reconstruction.

In Section~\ref{sec:strategy} we review the current observing
strategy and discuss its possible extensions.  In
Section~\ref{sec:sims} we briefly outline our numerical
simulation setting.  Section ~\ref{sec:CMBsims} describes the
construction of CMB templates from the currently available {\em
Planck} data.  Section~\ref{sec:sample} describes simulated
samples of galaxy clusters used for the flux-density
analyses. The main results are in Sec.~\ref{sec:results}.  Final
remarks and conclusions are in Sections~\ref{sec:discussion} and
~\ref{sec:conclusions} respectively.

\section{Observational strategy}
\label{sec:strategy}

\begin{figure*}[!t]
\centering
\includegraphics[width=0.49\textwidth]{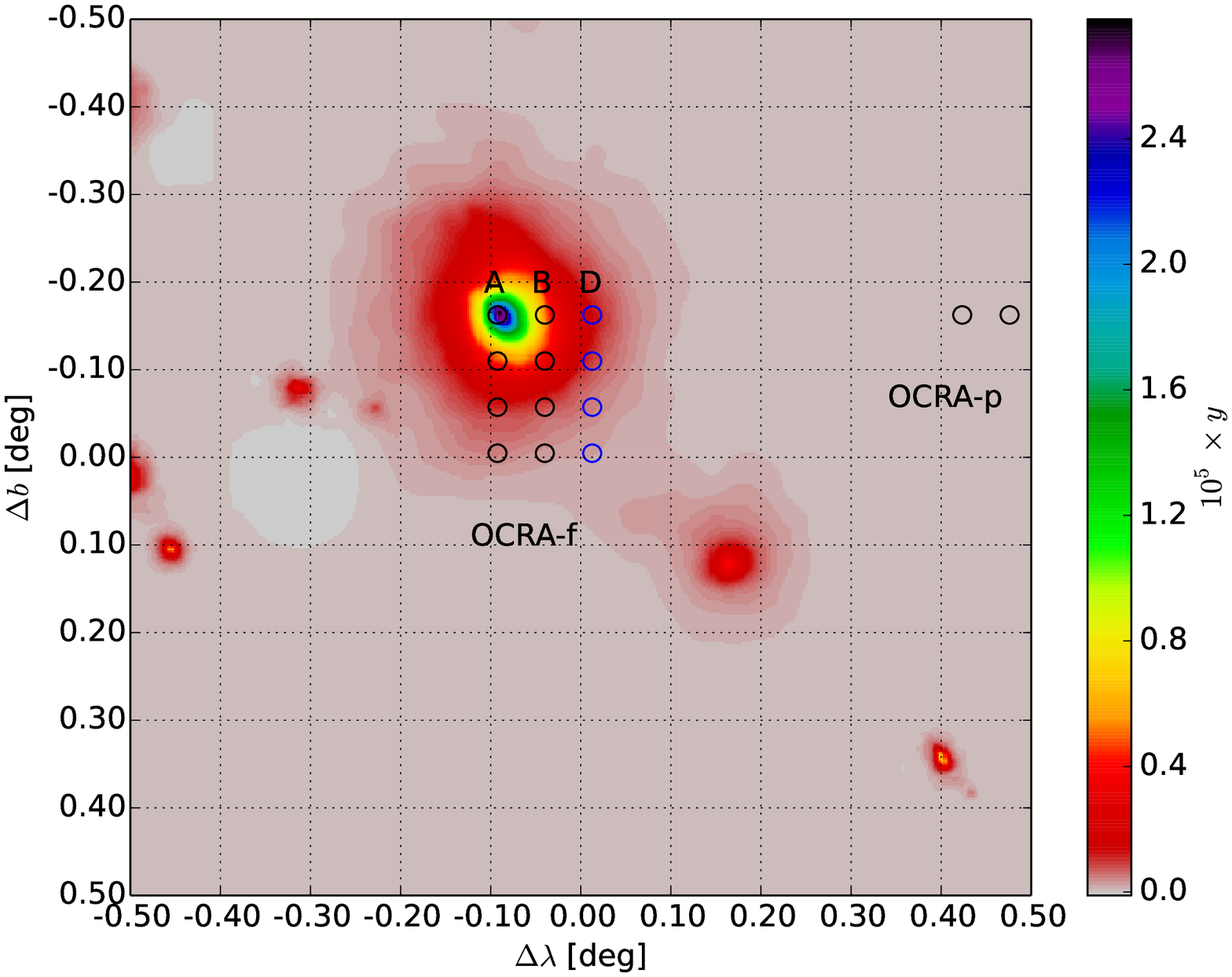}
\includegraphics[width=0.49\textwidth]{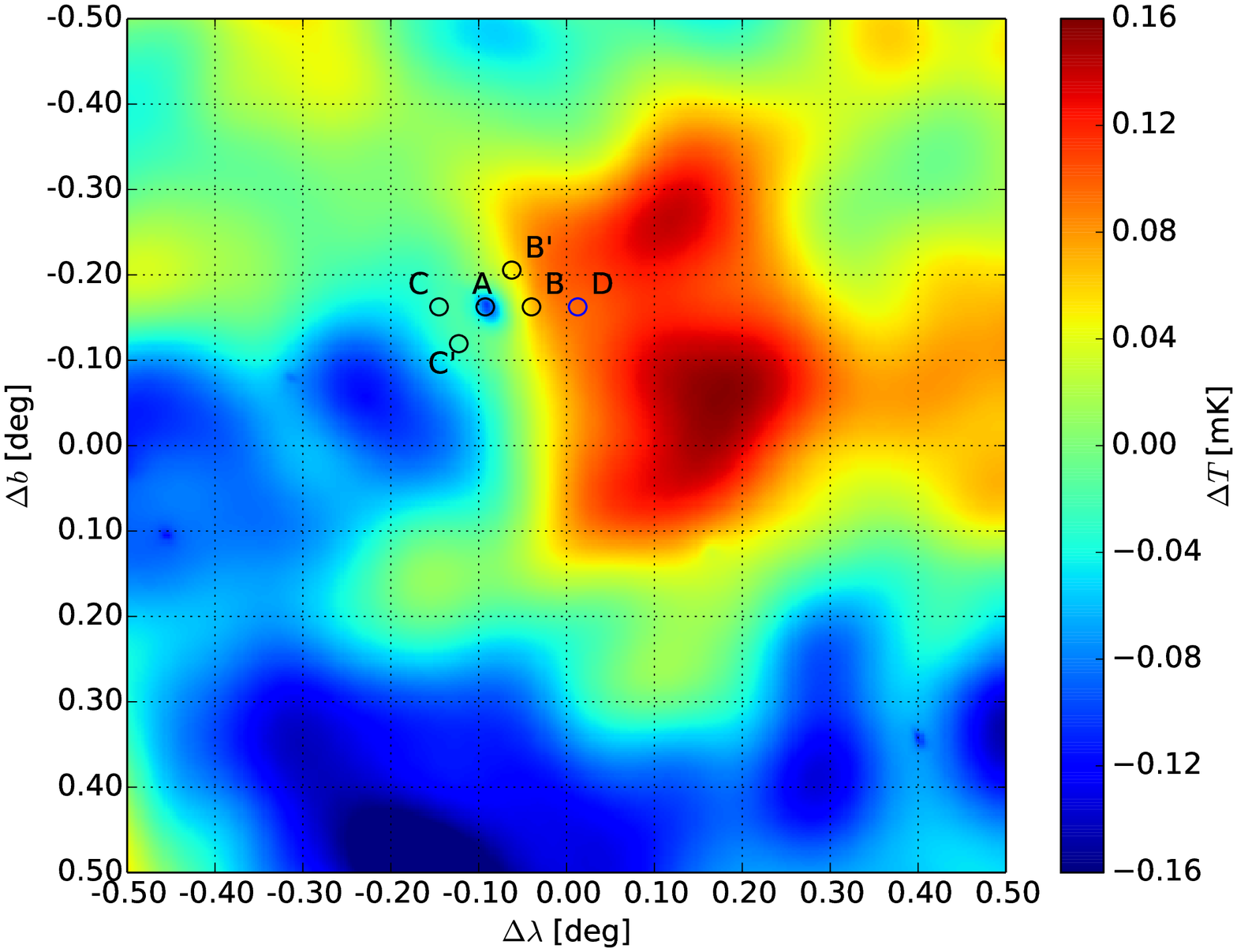}
\includegraphics[width=0.49\textwidth]{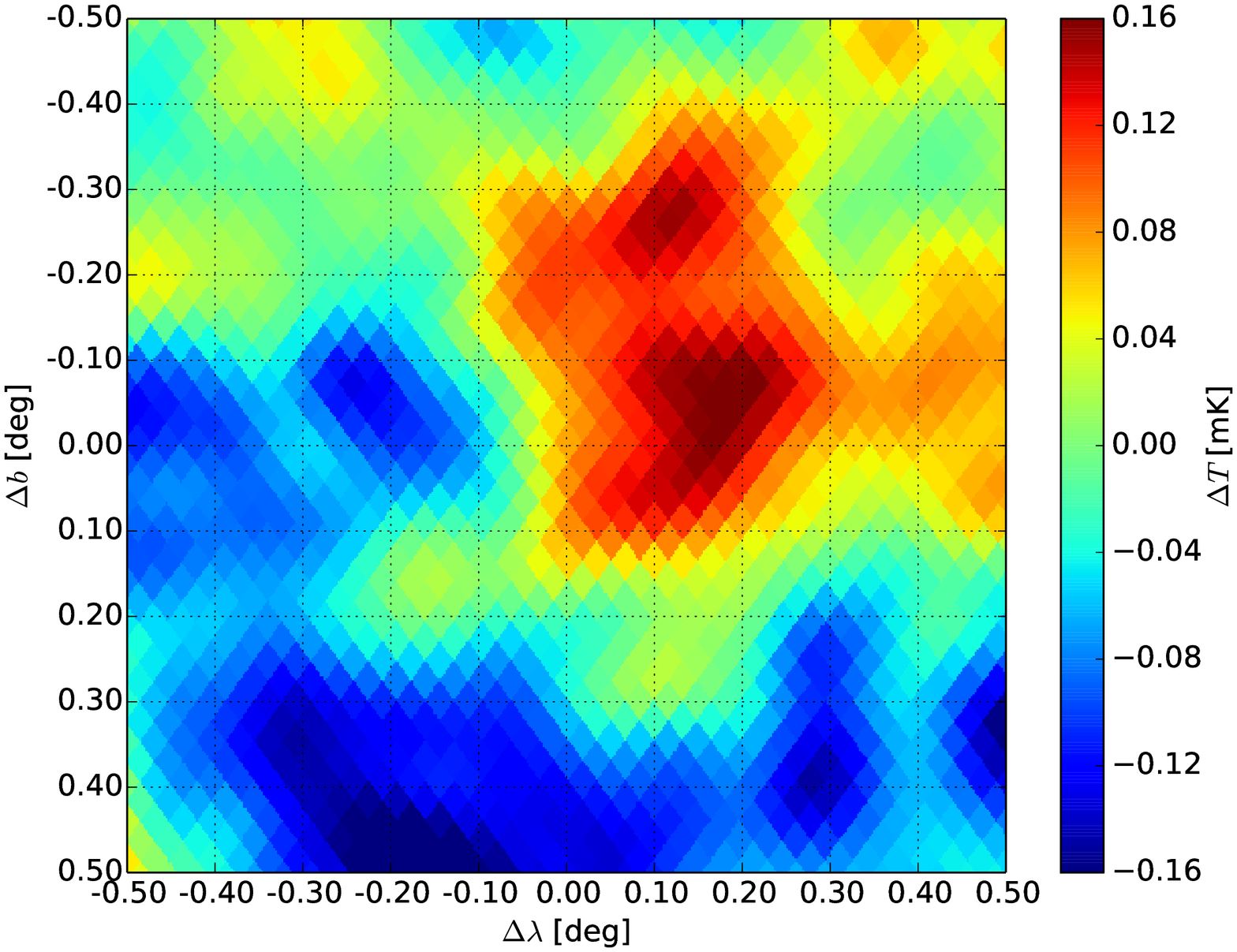}
\includegraphics[width=0.49\textwidth]{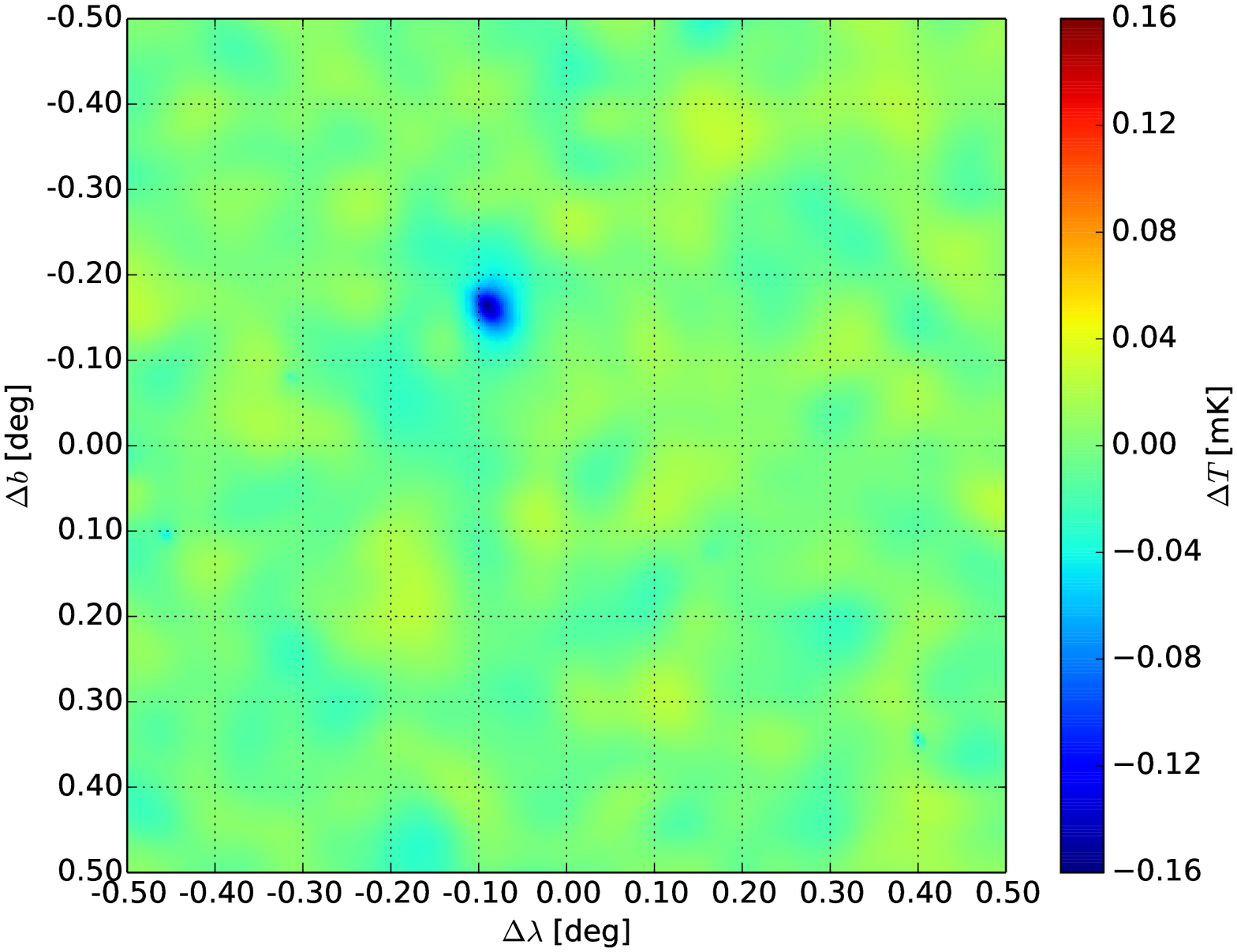}
\caption{ ({\em Top-left}) Projected map of the the line-of-sight
(LOS) integrated comptonization parameter for a selected halo,
and ({\em top-right}) simulated CMB temperature fluctuations
$\Delta T$ including both primary CMB as well as tSZ and kSZ
effects induced by the halo.  ({\em Bottom-left}) 217~GHz {\em
Planck} resolution simulation -- the Gaussian CMB $\Delta T$
signal smoothed with a half-power beam width (HPBW) of $\approx
5'$ and including a realistic {\em Planck} noise realization
(see Sec.~\ref{sec:CMBsims}). This map is used to make a high
resolution CMB template by means of a smooth-particle
interpolation.  ({\em Bottom-right}) Residual map: primary CMB
including tSZ minus interpolated {\em Planck} simulation
(Sec.~\ref{sec:CMBsims}).  The variance of the residual map
(without SZ effects) is about one order of magnitude smaller
than that measured using maps that include CMB.  In the {\em
top-right} panel the circles denote OCRA beamwidths
traversing over the background CMB for a typical observational
scheme (see Sec.~\ref{sec:strategy} for details).  In the {\em
top-left} panel black circles represent the full focal plane
of OCRA receivers.  }
\label{fig:templates}
\end{figure*}

The common position-switching mode of observing
\citep{Lancaster2011} is that in which the reference beam
non-uniformly (due to varying FOV rotation speed) integrates the
background along arcs $\approx 3'$ from the source.  In
Fig.~\ref{fig:templates} ({\em top-right} panel), circles denote
OCRA beamwidths for a typical observational scheme
\citep{Birkinshaw2005}.  First, the beam pair ``A--B'' measures
the difference signal between the cluster center and periphery,
respectively.  The beam pair is then ``switched'',
i.e. translated to configuration ``C--A'' with a swap of the
roles of the primary and reference beams, so that beams ``C'' and
``A'' now trace the cluster periphery and center, respectively.
The position switching cycle is closed by returning to the
initial configuration ``A--B'' and the cycle is repeated.

As the Earth rotates, the reference beams sweep arcs around the
cluster center and probe different off-center background regions
(beam B becomes B' and C becomes C').  The beam
position-switching reduces fluctuations due to atmospheric
turbulence on time scales of a few tens of seconds.  At shorter
time scales fluctuations due to receiver-gain instability and
atmospheric absorption are reduced by switching and
differentiating signals in receiver arms by means of
electronically-controlled phase switches
\citep{Peel2010,Lancaster2011}.\footnote{Beam switching is
realized at the rate of 277~Hz which improves the $1/f$ knee of
the resulting difference signal power spectrum roughly by an
order of magnitude; typically down to frequencies $0.1{\rm
Hz}<f_{\rm knee}<1{\rm Hz}$.}

An extension to this pattern can be realized by adding an extra
beam pointing ``D''.  In this case an observation cycle would be
extended so that a beam pair would observe differences between
``A--B'', then ``B--D'' and followed by ``A--B'', again probing
the cluster peripheries for varying parallactic
angles.  This scheme can be extended to both sides of the cluster
and to larger distances from the cluster center.  In
Fig.~\ref{fig:templates} ({\em top-left} panel) the projected
beamwidths of the OCRA-f receiver focal plane are shown (black
circles) overlaid on a nearby galaxy cluster seen through the SZ
effect. The extra beam pointing ``D'' for the whole array is
shown with blue circles.  In the following sections we
investigate the implications of such an extended observation
scheme using numerical simulations.

\section{Simulations}
\label{sec:sims}

\subsection{LSS and SZ effect}
\label{sec:LSSsims}

For the main results in this work we use the simulation approach
described in \cite{Lew2015}, with a few modifications.  In
particular, for the same field of view ($\approx 5.2^\circ$) we use
an increased map resolution of $\approx 0.9''$.  In observational
practise, often only the central comptonization parameter value
is quoted, so we include the kSZ signal calculated as a
contribution to the measured Compton $y$-parameter for the
appropriate frequency.

In this analysis we neglect the large-scale foreground
galactic synchrotron, free--free and dust emissions.  We 
assume these to be smooth enough to be 
removed in differential observations.

We assume that atmospheric effects and receiver noise are
mitigated by sufficiently long integrations (see
Sect.~\ref{sec:strategy} and \cite{Birkinshaw2005}).
Systematic errors in
flux density estimation that we neglect include feed and elevation dependent beam
response, feed and elevation dependent sidelobes, and elevation
dependent antenna gain.  We defer treatment of these effects to a
separate analysis of end-to-end full OCRA focal plane
simulations.  Although the simulated maps include some of the
effects of halo--halo LOS projections, the significance of the
projection effects on the $y$-parameter photometry are not the
main focus of the present study.  These
effects should be small in this study, since
we only consider the most massive systems at the
FOV generation stage.

A cluster is not necessarily observed at the position that
maximises the SZ signal.  For an X-ray selected galaxy cluster,
SZ observations can be centered at the maximum of the X-ray
signal, but this may have a small offset with respect to the
maximum of the SZ signal, although the difference should be
rather small.  This will effectively translate into pointing
errors that we model in this work.

The OCRA beam separation is $s_{\rm OCRA} = 0.0526\pm
0.0010^\circ$ (Fig.~\ref{fig:templates}), which we adopt in this
work.  Within the measurement accuracy this does not depend on
sky direction.

\subsection{CMB templates}
\label{sec:CMBsims}

The {\em Planck} mission has provided full sky maps of the CMB
temperature fluctuations at several frequencies, including
217~GHz, where the tSZ signal is minimal.  This opens up the
possibility to correct the differential SZ observations at other
frequencies by removing a customized CMB template and thus reduce
the CMB confusion in single frequency observations, provided
that: (i) the small scale noise level of the template is low,
(ii) the angular resolution of the data is sufficient and (iii)
diffuse Galactic foregrounds can be neglected.  In practise, 
requirement (iii) may exclude all or most of the sky regions
covered by {\em Planck} frequency maps on either side of the
Galactic Plane.

In order to test the usefulness of {\em Planck} data in
minimizing the CMB-induced variance in the measured flux
densities with a single-frequency instrument, we generate CMB
simulations for {\em Planck}'s 217~GHz frequency band using
simulated CMB power spectrum \citep{Lewis2000} and assuming
cosmological parameters as in \cite{Lew2015}.  We assume a
Gaussian beam transfer function defined by a full-width half
maximum FWHM~=~$5.02'$ \citep{PlanckCollaboration2016b} and use
realistic and publicly available 217~GHz {\em Planck} receiver
noise
simulations\footnote{\protect{\url{http://pla.esac.esa.int/pla/\#maps}}}
at Healpix \citep{Gorski2005} resolution $n_s=2048$.

We simulate the primary CMB up to $\ell_{\rm
max}=3500$,\footnote{The variance lost due to neglecting even
higher multipoles is negligibly small ($\lesssim 0.004\%$) and
well below the cosmic variance uncertainty.}  i.e. the OCRA
beam separation.  In order to create the final CMB template with
$\approx 0.9''$ resolution in small FOVs, we use a
smooth-particle interpolation of the projected field (as
discussed in \citealt{Lew2015}).

Real observations at 217~GHz will contain kSZ contributions,
which we ignore for generation of simulated templates.  For each
realization of the simulated template, we store maps with CMB
signal, and maps with the CMB smoothed with the instrumental beam
and contributed by a {\em Planck} noise realization
(Fig.~\ref{fig:templates}).  The former is used for simulating
astrophysical signals (SZ) in FOVs, while the latter is used for
removing a {\em Planck}-compatible version of CMB contamination,
for the same FOVs.  Each FOV simulation is made at fixed galactic
latitude $b=40^\circ$, but at a different galactic longitudes to
account for variations due to direction dependent properties of
{\em Planck} noise.

The single frequency maps from {\em Planck} mission are
contaminated by foregrounds other than the cluster tSZ and kSZ
signals, especially at low galactic latitudes.  Since in the
current work we do not investigate galactic foregrounds we
additionally analyze another set of simulations based on the
foreground-reduced map generated with the needlets-based internal
linear combination (NILC) algorithm \citep{PlanckCollaboration2016a}.  The
algorithm \citep{Delabrouille2009} provides a very clean primary
CMB map outside of masked regions which mask Galactic Plane and
bright point sources (together $\approx 3.6\%$ of the full sky).
We simulate the NILC map using the published beam transfer
function, and use the full mission NILC rendition of ringhalf-1
and ringhalf-2 half difference maps to generate a {\em Planck}
NILC-compatible noise realization.  We verified that the
resulting NILC simulations are compatible with the {\em Planck}
NILC map in terms of their angular pseudo-power spectra.  Slight
differences in the high-$\ell$ regime of up to a few percent are
present due to our choice of cosmological parameter values that
is consistent with the WMAP9 results.

The NILC map resolution simulations are similar to those of the
217~GHz map, so in Fig.~\ref{fig:templates} we only show the map
for the 217~GHz case.

\begin{figure}[!t]
\centering
\includegraphics[width=0.49\textwidth]{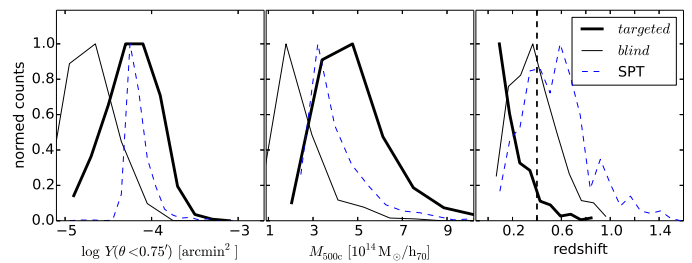}
\caption{ Distribution of the solid-angle--integrated
comptonization parameter $Y$ integrated within the angular
radius $\theta<0.75'$ from the cluster center ({\em left});
distribution of cluster total mass measured within a volume
with mean mass density $500$ times higher than the critical
density of the Universe at the cluster's redshift ({\em
middle}); and distribution of redshifts ({\em right}) for the
simulated cluster samples (solid lines) and for the SPT cluster
sample \citep{Bleem2015} (dashed lines).  For each simulated
sample only halos with $M_{\rm vir}>M_{\rm vir,\min}$ and
$z>z_{\min}$ are chosen (see Table~\ref{tab:samples}).  The
vertical dashed line shows the division into high-$z$ and
low-$z$ sub-samples that is used later in the analysis.  }
\label{fig:sample}
\end{figure}

\subsection{Simulated galaxy cluster samples}
\label{sec:sample}

\begin{table}[t]
\caption{Selection criteria used for constructing galaxy cluster samples.
}
\Beginruledtabular
\begin{tabular}{lcccc}
\BeginruledtabularSec
Parameter & \multicolumn{2}{c}{Sample/Value}\\
& \sampleHS & \sampleDF \\
$z_{\min}$ & 0.05 & 0.0 \\
$M_{\rm vir,\min}\,[10^{14} M_\odot/h]$ & 4.0 & 2.0 \\
halo selection & full simulation volume& $5.2^\circ\times 5.2^\circ$ FOV \\
halo count\footnotemark[1] & 475 & 361 \\
Sub-samples &&&& \\
\multicolumn{1}{r}{low-$z\leq0.4$} & 426 & 214  \\
\multicolumn{1}{r}{high-$z>0.4$}   & 49  & 147
\EndruledtabularSec
\footnotetext[1]{~The actual number of halos
used in statistical analyses are slightly different
as they are further screened for halos that lie well within the
projected FOV, which is required for simulating dual-beam
observations at all possible parallactic angles and beam separations in a consistent way.}
\end{tabular}
\Endruledtabular
\label{tab:samples}
\end{table}

For the analyses presented in Sect.~\ref{sec:results}, we
construct two galaxy cluster samples. The first one, hereafter
referred as \sampleHS (Fig.~\ref{fig:sample} thick solid lines)
is constructed by selecting the heaviest halos [$M_{\rm vir}>
4\times 10^{14} M_\odot/h$ (see Table~\ref{tab:samples})], from
each independent simulation volume and using each recorded
simulation snapshot.  We impose a low redshift cut-off
$z>z_{\min}$ to remove very extended clusters.  The choice of
redshifts for which simulation snapshots are taken is made such
that the simulation volume continuously fills comoving space out
to the maximal redshift (see Fig.1 of~\cite{Lew2015}).  For each
simulation volume we apply random periodic coordinate shifts of
the particles within, and we apply random coordinate switches.
This (i) improves redshift space coverage and (ii) yields cluster
SZ surface brightness profiles in different projections, at the
cost of generating a partially correlated sample.

The second sample, hereafter referred to as \sampleDF
(Fig.~\ref{fig:sample} thin solid lines), is generated using a
blind survey approach (as in ~\cite{Lew2015}). We generate 37 FOV
realizations each $\approx 27\,{\rm deg}^2$ together covering a sky
area of $\approx 1000\, {\rm deg}^2$.  From each realization we
select halos with virial masses $M_{\rm vir,c} > 2\times 10^{14}
M_\odot/h$.

The solid angle integrated comptonization for any given halo
(${\protect Y=\int y(\hat {\mathbf n}) d\Omega}$) depends on a
combination of halo redshift and mass.  The \sampleDF sample is
dominated by lighter halos than those found in the SPT sample
(Fig.~\ref{fig:sample}), although redshift space distributions of
the two are similar. Hence, the bulk of the \sampleDF sample
halos yields lower $Y(\theta<0.75')$ values than those in the SPT
sample (Fig.~\ref{fig:sample}).  Although increasing the lowest
mass limit for the halos of the \sampleDF sample tends to make
its mass and redshift distributions more consistent with those of
the SPT sample, it reduces the numbers of halos, thus increasing
Poisson noise.  For the statistical analysis in
this work, larger simulations and more FOV realizations than are
currently available would be required to reach
consistency. Therefore, we use this sample for tSZ analyses of
simulated dual beam observations, bearing in mind that in this
limit of weak SZ effects, CMB confusion is expected to be the
most significant. On the other hand, the \sampleHS sample is
expected to be less affected by CMB confusion.

In order to investigate the differences between compact and
extended SZ clusters we further split our cluster samples by
redshift at $z=0.4$ (Table~\ref{tab:samples}).  This split roughly
corresponds to half of the radial comoving distance to $z=1.0$,
beyond which we do not observe any heavy (Fig.~\ref{fig:sample})
halos in our simulations.
We find that the low-$z$ and high-$z$
samples mainly differ due to the strength of the SZ effects,
and due to the presence of sub-structures,
being respectively stronger and more abundant in the low-$z$ subset.
Examples for halos from low-$z$ and high-$z$ samples are shown
below in Fig.~\ref{fig:halos}.

\begin{figure*}[!t]
\centering
\includegraphics[width=0.8\textwidth]{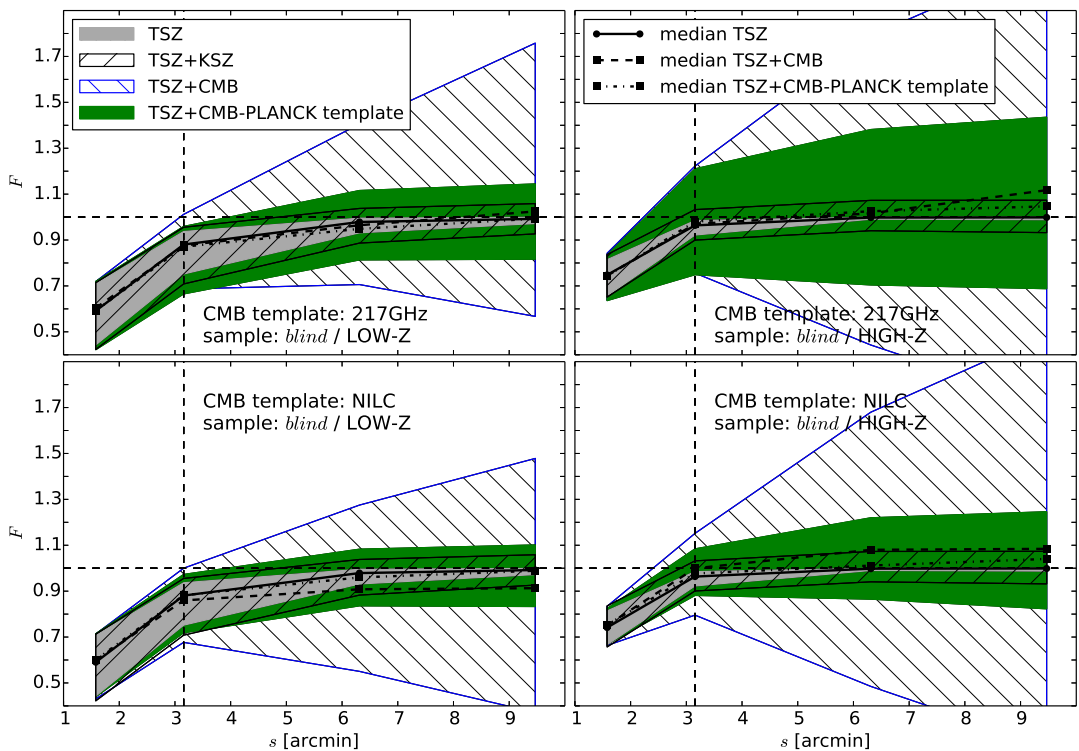}
\caption{Simulated fractions ($F$) of SZ effect flux density at
30~GHz recovered from difference, dual beam observations of
clusters from \sampleDF sample as a function of beam angular
separation $s$ and redshift range.  The shaded/hatched regions
map the 68\% confidence regions (CR) in the distribution of
$F$.  The scatter in $F$ calculated from maps containing only
tSZ signal is shown in gray.  The backslash-hatched region
shows the effects of primary CMB on biasing the tSZ flux
density measurements.  The forward-slash--hatched region shows
the intrinsic scatter due to kSZ when converted and embedded
into the 30~GHz thermal SZ effect maps.  The green region shows
the improvements in decreasing the intrinsic scatter in $F$ as
a result of subtracting the Planck CMB template from CMB+tSZ
simulated maps prior to flux density calculations.  The median
$F$ values are shown as lines.  The 68\% confidence regions
about the medians become asymmetric as the beam separation
increases (simulation sample error also becomes obvious in the
TSZ+CMB case by comparing upper to lower plots; the TSZ,
TSZ+KSZ, and TSZ+CMB cases are statistically equivalent between
the upper and lower panels).  The vertical dashed line marks
the actual separation of OCRA beams fixed by the telescope
optics.  It is assumed that the reference beam covers an
annulus around a galaxy cluster within parallactic angle range
$[0^\circ,180^\circ]$ on either side of the central direction,
and that pointing error $\epsilon_p=0$ (see
Sec.~\ref{sec:pointing}).  }
\label{fig:medianfDF}
\end{figure*}

\begin{figure*}[!t]
\centering
\includegraphics[width=0.8\textwidth]{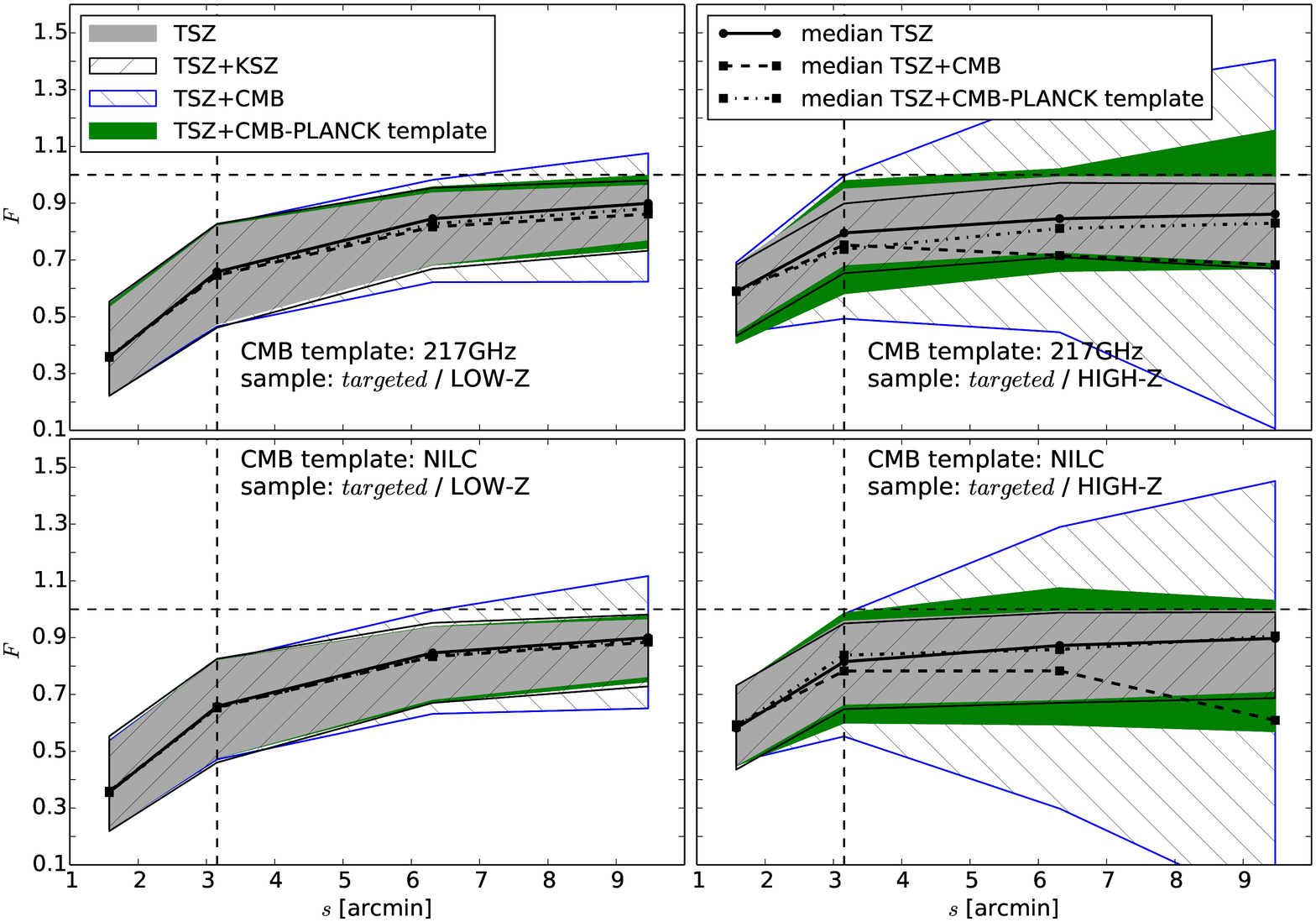}
\caption{As in Fig.~\ref{fig:medianfDF} but for the \sampleHS sample.}
\label{fig:medianfHS}
\end{figure*}

\section{Analysis and results}
\label{sec:results}

\subsection{Systematic effects from beam separation}
\label{sec:beam_separation}

Dual beam difference observations capture only a fraction of the
intrinsic flux density, depending on the physical extent of the
source, its redshift and the angular separation of the beams.  We
define this fraction as
\begin{equation}
\label{eq:F}
F(\theta_b, s,q_{\rm max}) = \bigg\langle \frac{S_0^{\rm x}(\theta_b,s) - S_r^{\rm x}(\theta_b,s,\mathbf{\hat n_i})}{S_0^{\rm tSZ}(\theta_b)} \bigg\rangle_i,
\end{equation}
where $S_0^{\rm x}$ is the measured central flux density per beam
induced by effect ``x'', e.g. x~=~tSZ, $S_0^{\rm tSZ}$ is the
true central flux density per beam due to tSZ (neglecting CMB,
point sources, and other effects), $S_{\mathrm r}^{\rm x}$ is the
flux density per beam due to effect ``x'' in the reference beam
direction ($\mathbf{\hat n}_i$), $\theta_b$ is the instrumental
half-power beam width (HPBW), and $s$ is the angular separation
of the beams.  $N_{\mathrm r}=500$ reference beam directions
($\mathbf{\hat n}_i$) are chosen randomly from a uniform
distribution of parallactic angles ($q \in [0,q_{\max}]$), where
the upper limit $q_{\max}$ is a free parameter.

For each halo, we measure these fractions $F$ by integrating
specific intensity directly from high resolution maps, and using
the mean over the $N_{\mathrm r}$ values of $q$.  If there is no
CMB contamination, i.e. setting x~=~tSZ, so that
$S_0^{\rm x} = S_0^{\rm tSZ}$
and
$S_{\mathrm r}^{\rm x} = S_{\mathrm r}^{\rm tSZ}$,
then $F\leq 1$ and $1-F$ represents the fraction of the signal
lost only due to the closeness of the beam angular separation.

The impact of beam separation on the dual-beam observations is
shown in Figs.~\ref{fig:medianfDF} and~\ref{fig:medianfHS} for
the \sampleDF and \sampleHS samples respectively, for an
idealistic case of exact pointing---i.e., no pointing
inaccuracies are allowed ($\epsilon_p=0$).  In these figures, the
median $F$ (from all halos matching the selection criteria) is
plotted along with a 68\% confidence region.  Clearly, dual-beam
observations at larger beam separations are less biased than
observations at smaller beam separations, and the 68\% confidence
range generally shrinks as $s$ increases.

The significance of the primary CMB fluctuations for the
dual-beam observations is estimated by setting x~= tSZ$+$CMB,
i.e.,
$S_0^{\mathrm x}=S_0^{\mathrm{tSZ+CMB}}$, 
$S_{\mathrm r}^{\mathrm x}=S_{\mathrm r}^{\mathrm{tSZ+CMB}}$.
While the median $F$ does not differ significantly from the pure
tSZ case, the 68\% confidence region significantly increases with
beam separations due to primary CMB confusion.  For example,
since a primordial CMB fluctuation has a good chance of being of
the same sign as the SZ signal at the cluster center but of the
opposite sign in a distant reference beam, $F$ can easily be
greater than unity, as is clear in Fig.~\ref{fig:medianfDF}.  As
expected, the increase is stronger in the \sampleDF
sample/high-$z$ sub-sample than in the \sampleHS sample/low-$z$
sub-sample, due to differences in amplitudes of SZ effects
compared to the level of CMB fluctuations.

Comparing Figs.~\ref{fig:medianfDF} and ~\ref{fig:medianfHS} it
is clear that the main difference is the relative significance of
the CMB as a source of confusion and the amount of residual
biasing.  However, for any individual high-$z$ and/or low-mass
cluster observation, the measured flux density can be biased
substantially. 
This can be inferred from the size of the
$1\sigma$ tSZ+CMB confidence region. Even observations of the
most massive clusters, which are the least affected by the
presence of the CMB, can be biased substantially depending on the
angular scales being measured ($s$) (Fig.~\ref{fig:medianfHS}
left panels).  In the figure, the trade-off
between CMB confusion due to observations at larger angular
scales and the level of biasing ($F$) in the limit of small $s$
is clearly seen.

For clusters that are small relative to the beam size,
measurements far away from the cluster center are not really
needed as the $F$ values approach unity relatively fast
(e.g. \sampleDF/ high-$z$ sample in Fig.~\ref{fig:medianfDF}).
At the OCRA beam separation (the vertical line in the figures)
the primary CMB does not strongly contribute to the scatter in
flux density measurements.  This is even more so in the case of
the \sampleHS sample of heavy and low-$z$ clusters.  On the other
hand, the most massive halos (Fig.~\ref{fig:medianfHS}) require
significant ($> 10\%$) flux density corrections even at
large beam separations (although these may partially be generated
by projection effects discussed in Sec.~\ref{sec:discussion}).

It is clear that in the two cluster samples, kSZ 
only slightly increases the scatter in $F$ at the OCRA beam
separation, as expected at 30~GHz.

The impact of {\em Planck} based CMB template removal is shown in green.
The calculation is done by setting x~= tSZ$+$CMB$-$template 
in Eq.~\ref{eq:F},
i.e.,
$S_0^{\mathrm x}=S_0^{\mathrm{tSZ+CMB-template}}$, 
$S_{\mathrm r}^{\mathrm x}=S_{\mathrm r}^{\mathrm{tSZ+CMB-template}}$.
From Figs.~\ref{fig:medianfDF} and~\ref{fig:medianfHS} it is
clear that at the OCRA beam separation, and for the full range of
parallactic angles, the {\em Planck} template does not
significantly help, or does not help at all, in reducing the
confusion due to primary CMB. However, in observations that probe
larger angular separations, the CMB template removal can
substantially reduce the $1\sigma$ contours. The template removal
may also be useful for observations of high-$z$ massive clusters
for which mapping larger angular distances away from the central
directions still appears to be well motivated. Both in the
high-$z$ and low-$z$ sub-samples of the \sampleHS sample the
template reduces the tSZ+CMB scatter nearly down to the level
limited by the intrinsic tSZ scatter for the full range of $s$
studied here (Fig.~\ref{fig:medianfHS}).

\subsection{Parallactic angle dependence}
\label{sec:PA}

\begin{figure}[!t]
\centering
\includegraphics[width=0.49\textwidth]{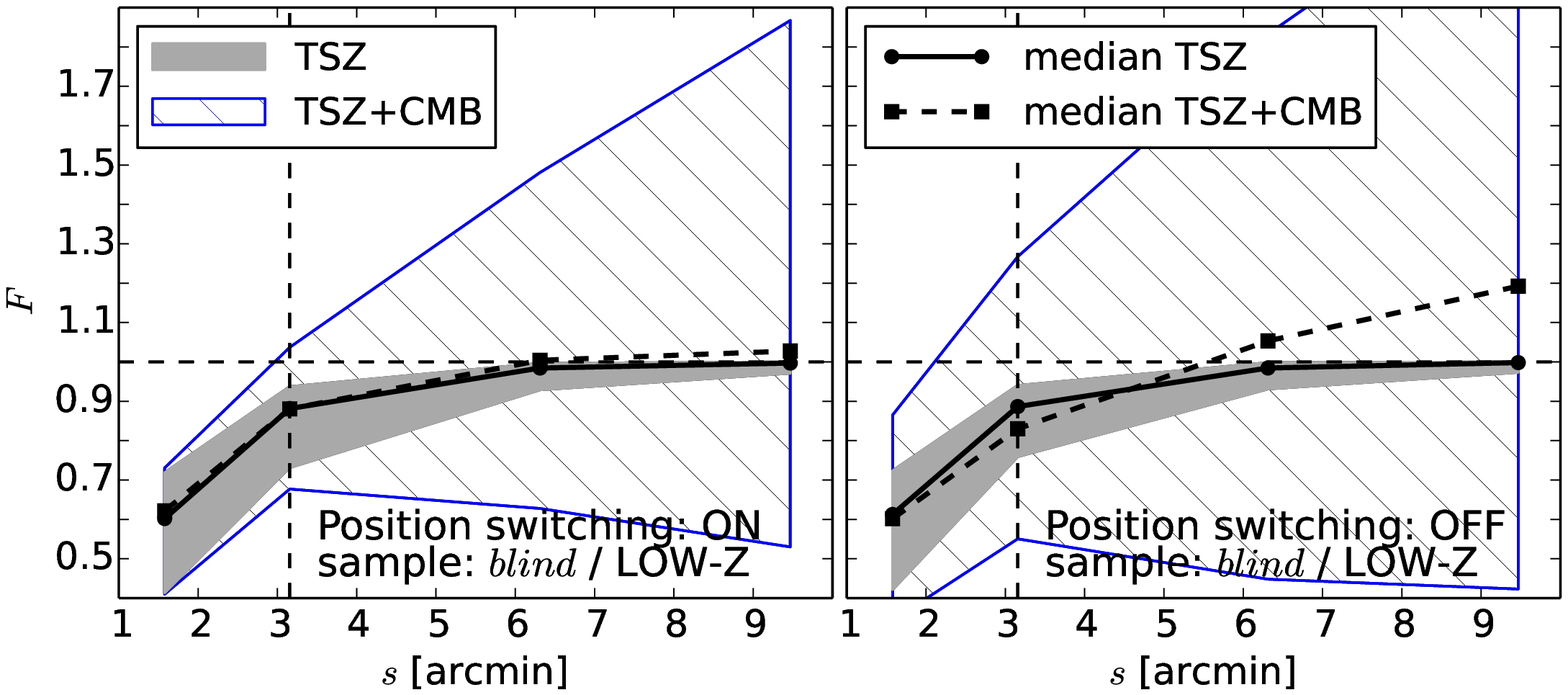}
\includegraphics[width=0.49\textwidth]{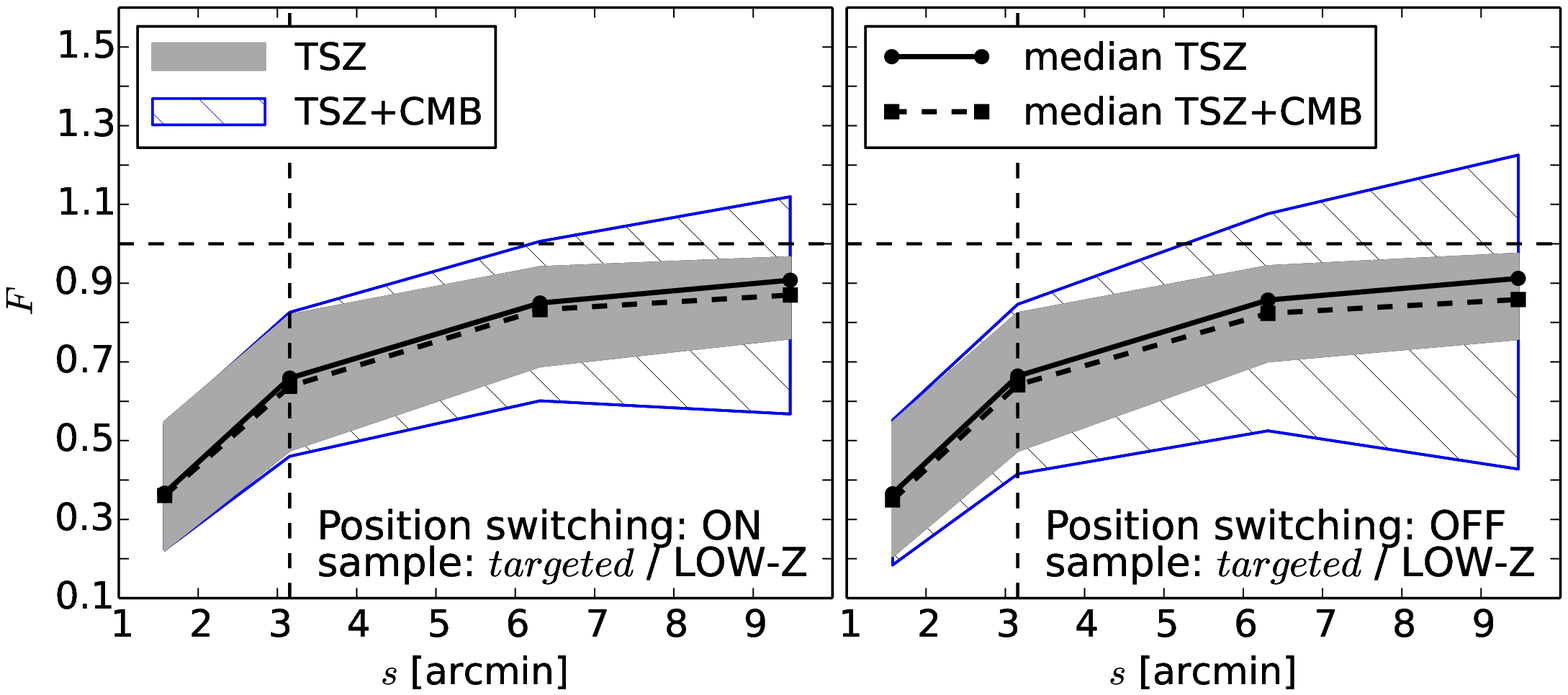}
\caption{As in Figs~\protect\ref{fig:medianfDF} and
\protect\ref{fig:medianfHS}, but for observations where the
reference beam covers an annulus around a galaxy cluster within
parallactic angle range $[0^\circ,22.5^\circ]$ on either side
of the cluster direction ({\rm left}), or only on one side of
the cluster direction ({\rm right}) for the \sampleDF (top) and
\sampleHS (bottom) samples.  }
\label{fig:F_vs_PA_DF}
\end{figure}
In practise, the OCRA observations exploit beam and position
switching (Sect.~\ref{sec:strategy}) but it is unrealistic to
cover the full parallactic angle range: i.e. $q\in
[0^\circ,q_{\max}]$ where $q_{\max}=360^\circ$.

It was already known that position switching
\citep{Birkinshaw2005} significantly mitigates atmospheric
instabilities over the time scale of tens of seconds by (i)
subtracting linear drifts caused by large-scale precipitable
water vapor (PWV) fluctuations \citep{Lew2016}, (ii) accounting
for beam response asymmetries, and (iii) maximizing the
probability of avoiding (masking out) intervening radio sources
that can significantly bias the SZ measurement.  In this section,
we show that position switching is also efficient in mitigating
the confusion due to primordial CMB, even with a very modest
coverage of parallactic angles.

There should not be any statistical correlation between
primordial CMB fluctuations and the locations of heavy
halos. Moreover, galaxy clusters have small angular sizes
compared those representing most of the CMB power.  Thus,
clusters should mostly lie on slopes rather than peaks or troughs
in the CMB map. Hence, sampling SZ flux density differences at
opposite sides of a galaxy cluster core should help average out
the primordial CMB in comparison to one-sided observations.  We
confirm that this is indeed the case and find that this
improvement is reached at even moderate values of $q_{\max}$.

We calculate $F(s)$ [Eq.~(\ref{eq:F})] for maps containing tSZ
and CMB using mean flux density estimates either according to the
position switching observation scheme or without it.  As before,
each measurement is an average of $500$ dual-beam pointings at
different $q$ but drawn randomly from within the range
$[0^\circ,q_{\max}]$ where
$q_{\max}\in\{180^\circ,90^\circ,45^\circ,22.5^\circ\}$.

The result is shown in Fig.~\ref{fig:F_vs_PA_DF} for
$q_{\max}=22.5^\circ$.  By comparing the left panel of this
figure with the top-left panel of Fig.~\ref{fig:medianfDF} it is
clear that even strongly incomplete coverage of the parallactic
angles does not cause significant broadening of the 68\%
confidence level (CL) contours.  However, when position switching
is not used (right panels in Fig.~\ref{fig:F_vs_PA_DF}), the
confusion due to primordial CMB is stronger. As before, the
\sampleHS sample of the heaviest halos is less affected by the
presence of the CMB, but the effect of not using position
switching is still visible, even at $s=s_{\rm OCRA}$ (vertical
line in Fig.~\ref{fig:F_vs_PA_DF}, bottom panels).

\subsection{Systematic effects in redshift space}
\label{sec:Fz}

\begin{figure*}[!t]
\centering
\includegraphics[width=0.49\textwidth]{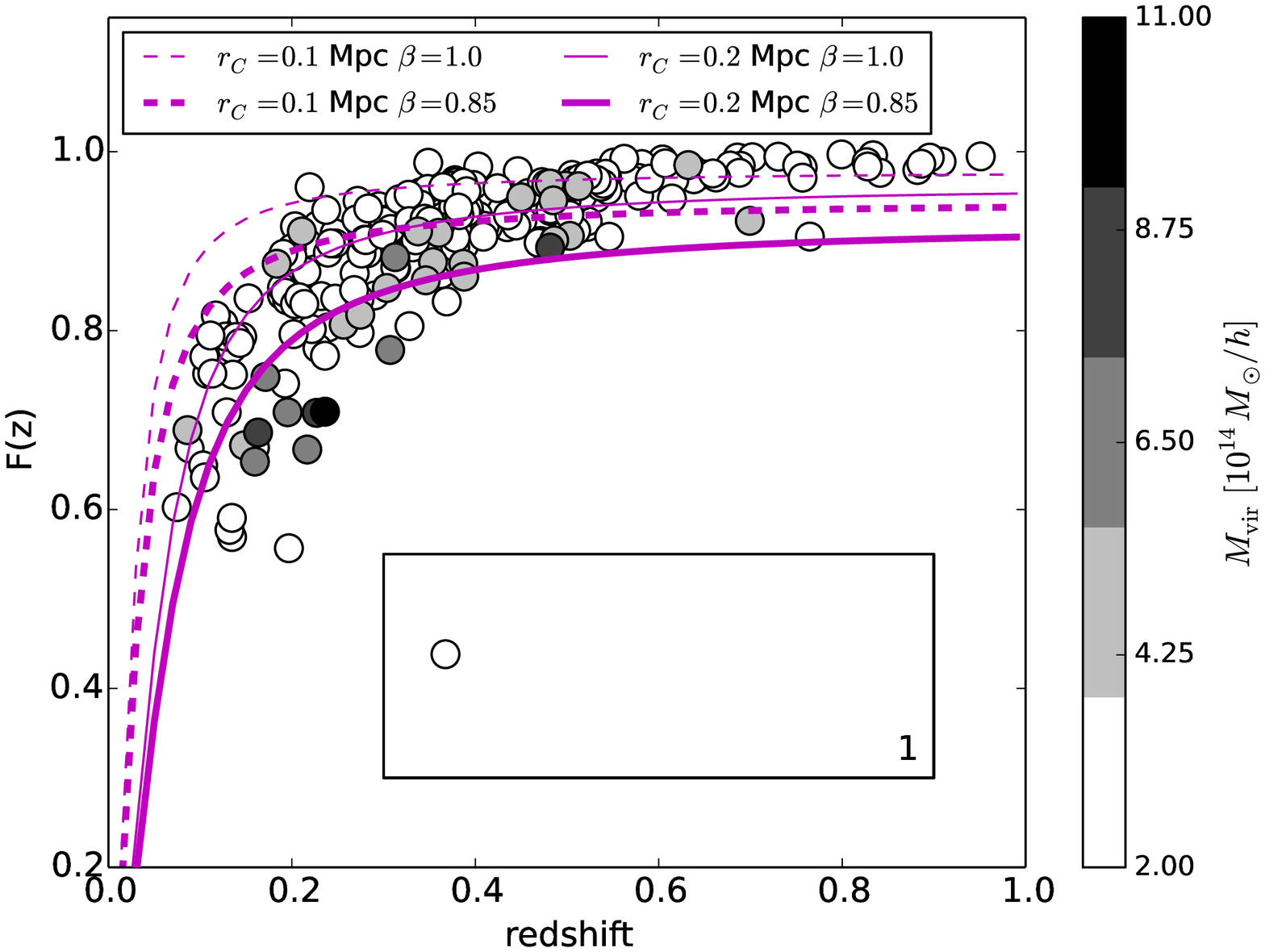}
\includegraphics[width=0.49\textwidth]{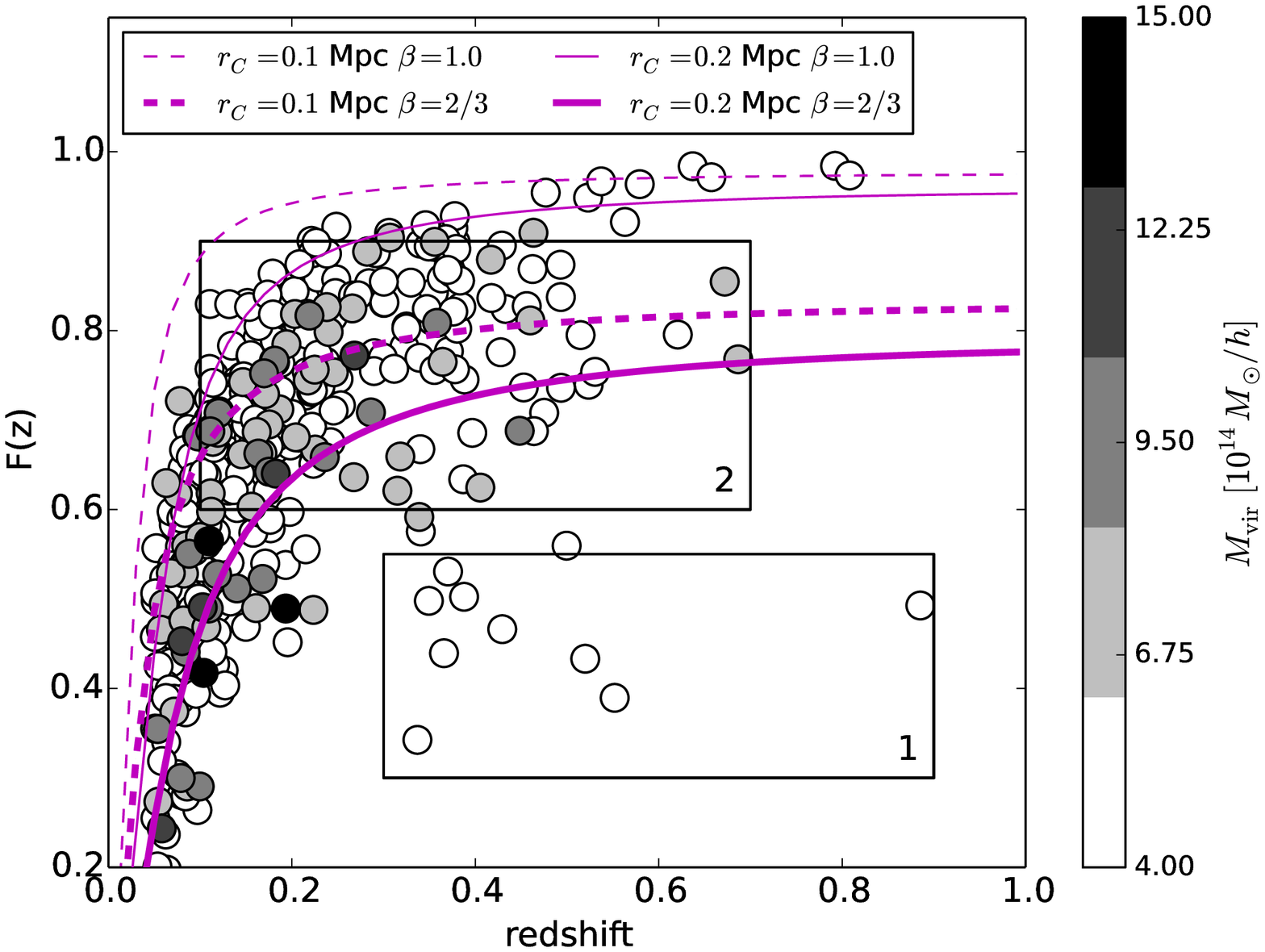}
\caption{Simulated fractions ($F$) as a function of redshift and
virial mass for clusters from \sampleDF (left) and \sampleHS
(right) samples and for observations at effective beams angular
separation ${\protect s=0.0526^\circ}$ (the separation of OCRA
beams).  The fractions were measured from maps containing tSZ
signal only.  The lines trace the dependence for a halo
described by a $\beta$-model according to Eq.~\ref{eq:Fz} with
parameters given in the plot legend.  The rectangles ``1'' and
``2'' mark strong outliers and some halos from the main group
that are inspected individually (see text for discussion).  }
\label{fig:f_z}
\end{figure*}

The $F$ factor depends on a cluster's angular size, which in turn
depends on the cluster's redshift.  We model the dependence of
$F(\theta_b, s,p)$ (Eq.~\ref{eq:F}) on redshift by defining:
\begin{equation}
\label{eq:Fz}
F_{\mathrm m}(\theta_b, s,\beta,\theta_c) =
1 - \frac{\int b(\mathbf{\hat n},\theta_b,s) \, I_\mathrm{SZ}(\mathbf{\hat n},\beta,\theta_c)\d\Omega}{\int b(\mathbf{\hat n},\theta_b,s=0) \, I_\mathrm{SZ}(\mathbf{\hat n},\beta,\theta_c)\d\Omega},
\end{equation}
where $b(\mathbf{\hat n},\theta_b,s)$ is a Gaussian beam profile
with beam width $\theta_b$ , offset by angular distance $s$ from
the cluster center direction $\mathbf{\hat n_0}$.  We choose
$s=s_{\rm OCRA}$ to simulate the position of the OCRA reference
beam when the primary beam points at the cluster center.
$I_\mathrm{SZ}(\mathbf{\hat n},\beta,\theta_c)$ is a normalized,
LOS integrated $\beta$ profile that represents the SZ effect
surface brightness:
\begin{equation}
\label{eq:beta}
I_\mathrm{SZ}(\mathbf{\hat n},\beta, \theta_c) \propto
\left(1+ \frac{\theta^2}{\theta_c^2} \right)^{\frac{1}{2}-\frac{3}{2}\beta},
\end{equation}
where $\theta_c=2 r_c/d_A(z)$ is the angular diameter of the
observed galaxy cluster defined in terms of its core size $r_c$,
$\theta$ is the angle from $\mathbf{\hat n}$ to $\mathbf{\hat
n_0}$, and $d_A(z)$ is the angular diameter distance.
$F_{\mathrm m}$ depends on the choice of cosmological parameters
and on the chosen cluster density profile. We calculate
$F_{\mathrm m}$ for $\Lambda$CDM cosmological parameters:
$h=0.7$, $\Omega_m=0.3$, $\Omega_\Lambda=0.7$, and for an
Einstein--de~Sitter cosmological model. For our redshift range we
find that the dependence on cosmological parameters is weak
compared to the dependence on the halo density profile
(Fig.~\ref{fig:f_z}).  We also calculate $F_{\mathrm m}$ for the
case of a Gaussian halo but find that such profile is strongly
disfavored by simulations as $F_{\mathrm m}$ approaches unity at
fairly low redshifts.

The simplest $\beta$-model does not allow for the steepening of
density profiles with increasing $\theta$.  However, X-ray
observations suggest that such steepening is real
(e.g. \citealt{Vikhlinin2006}), and it is expected that at large
distances from cluster cores (or higher redshifts) the
$\beta$-model yields lower $F_{\mathrm m}(z)$ values than those
predicted by simulations, as seen in Fig.~\ref{fig:f_z}.

Clearly, the \sampleHS sample has a large scatter in $F$ values
at high redshifts. Some of that scatter is due to projection
effects, which we discuss latter.  Heavy clusters of the
\sampleHS sample appear more compatible with the $\beta$-model at
lower $\beta$ values than the lower-mass clusters of the
\sampleDF sample.  At the OCRA beam separation, the low-mass
clusters in both samples show very weak effects of biasing
($F\approx 1$) at the highest redshifts.  On the other hand heavy
halos require large corrections, some of which do not result from
simple projection effects. In the next section, a selection of
halos are investigated individually.

\subsection{Analysis of individual clusters}

\begin{table}[t]
\caption{Parameters of halos selected from Fig.~\ref{fig:f_z}.
The parameter $y_0$ is the maximal value of the LOS integrated
comptonization parameter.  }
\Beginruledtabular
\begin{tabular}{lccccc}
\BeginruledtabularSec
ID & $z$ & $F$ & $M_{\rm vir}$ & $y_0\times 10^5$ & comment\footnotemark[1]\\
& & & $[10^{14} M_\odot/h]$  &&\\
\hline
\multicolumn{6}{c}{\sampleHS sample (rect. ``1'' selection)}\\
1  & 0.552 & 0.39 & 4.6  & 1.38 & P \\
2  & 0.520 & 0.43 & 4.7  & 3.40 & P \\
3  & 0.885 & 0.49 & 4.4  & 0.71 & P, D \\
4  & 0.429 & 0.47 & 4.6  & 2.10 & P, E \\
5  & 0.366 & 0.44 & 5.1  & 1.75 & P \\
6  & 0.370 & 0.53 & 4.9  & 2.10 & P \\
7  & 0.349 & 0.50 & 5.1  & 3.10 & P \\
8  & 0.337 & 0.34 & 5.4  & 3.70 & P \\
9  & 0.388 & 0.50 & 4.4  & 2.2  & P \\
\multicolumn{6}{c}{\sampleHS sample ($M_{\rm vir}>12.7\times 10^{14} M_\odot/h$)}\\
10 & 0.103 & 0.41 & 13.8 & 11.2 & D \\
11 & 0.109 & 0.56 & 14.6 & 25.6 & R, P \\
12 & 0.193 & 0.49 & 12.8 & 5.6  & E, D, S, P\\
\multicolumn{6}{c}{\sampleDF sample (rect. ``1'' selection)}\\
13 & 0.367 & 0.44 & 2.1  & 0.93  & P \\
\multicolumn{6}{c}{\sampleHS sample (rect. ``2'' selection)}\\
14 & 0.182 & 0.64 & 11.3 & 6.68  & D \\
15 & 0.268 & 0.77 & 12.6 & 9.50  & D \\
16 & 0.169 & 0.75 & 10.1 & 10.3  & R, S\\
\EndruledtabularSec
\end{tabular}
\Endruledtabular
\footnotetext[1]{P - reference beam flux density contamination
from another halo due to LOS projection; D - disturbed
morphology; E - elongated shape; R - regular morphology
(virialized halo); S - sub-halo(s) present;}
\label{tab:halos}
\end{table}

In Fig.~\ref{fig:f_z} some of the halos are selected by
rectangles in the $z-F$ diagram.  The properties of some of these
halos are given in Table.~\ref{tab:halos}.  Fig.~\ref{fig:f_z}
shows that only the lightest halos in our samples are found to be
strong outliers, which is unsurprising.

\begin{figure}[!t]
\centering
\includegraphics[width=\haloPlotWidth\textwidth]{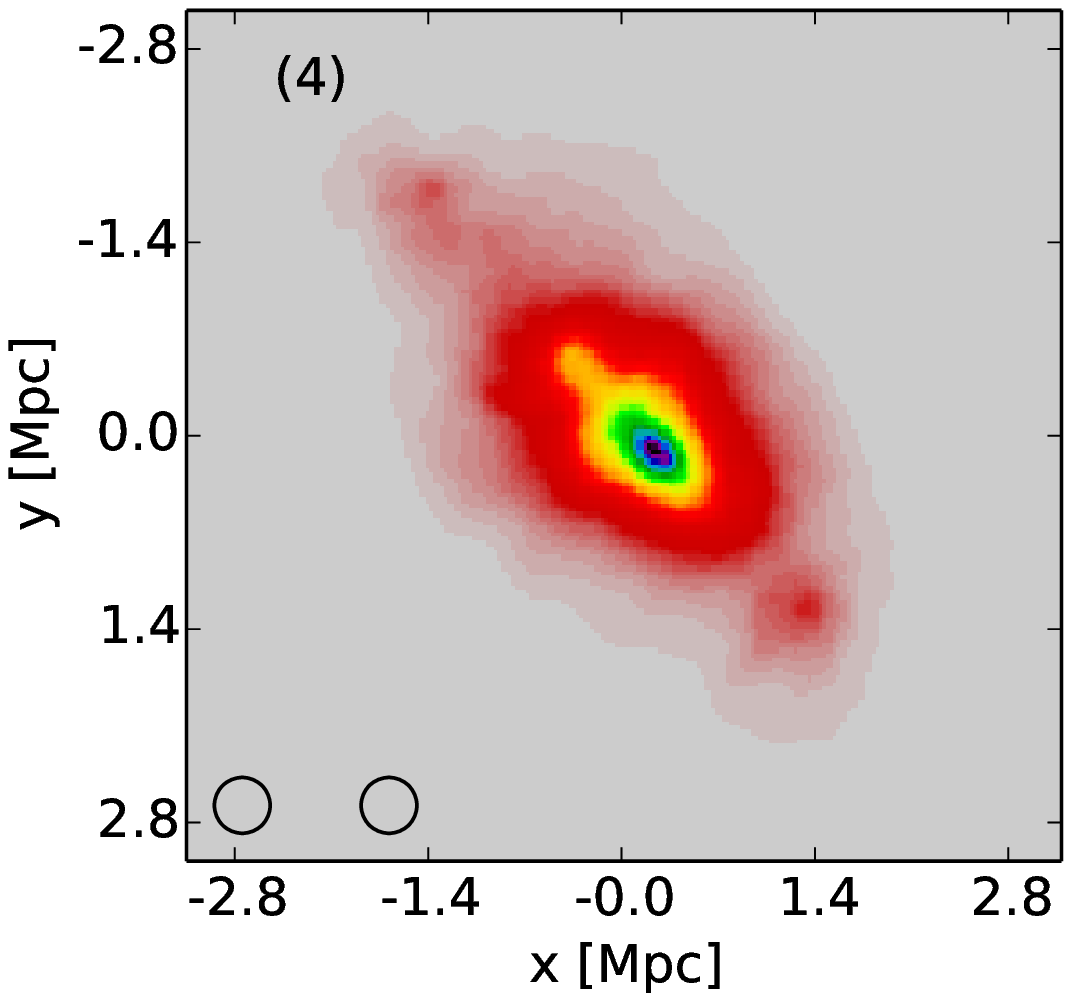}
\includegraphics[width=\haloPlotWidth\textwidth]{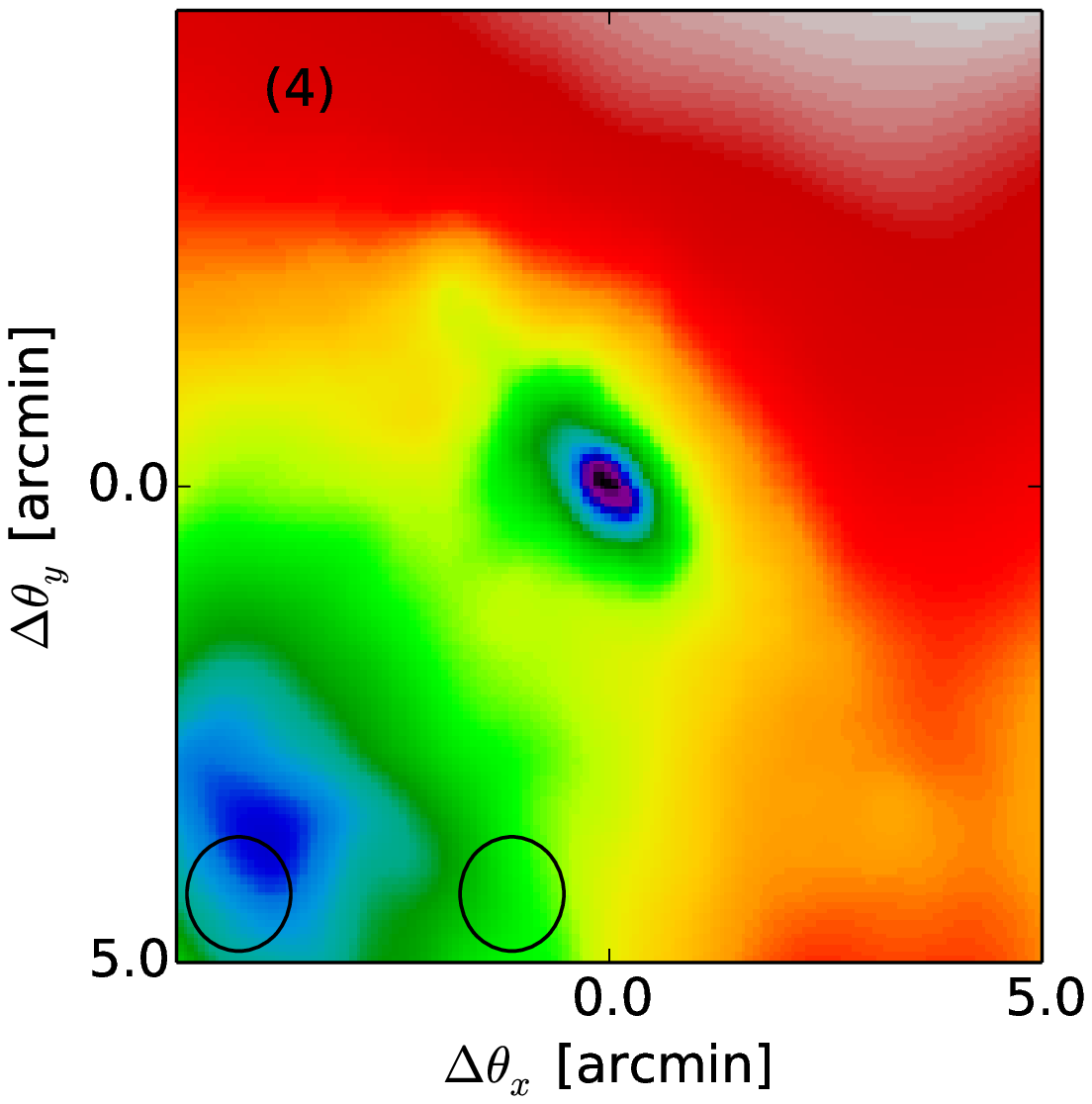}
\includegraphics[width=\haloPlotWidth\textwidth]{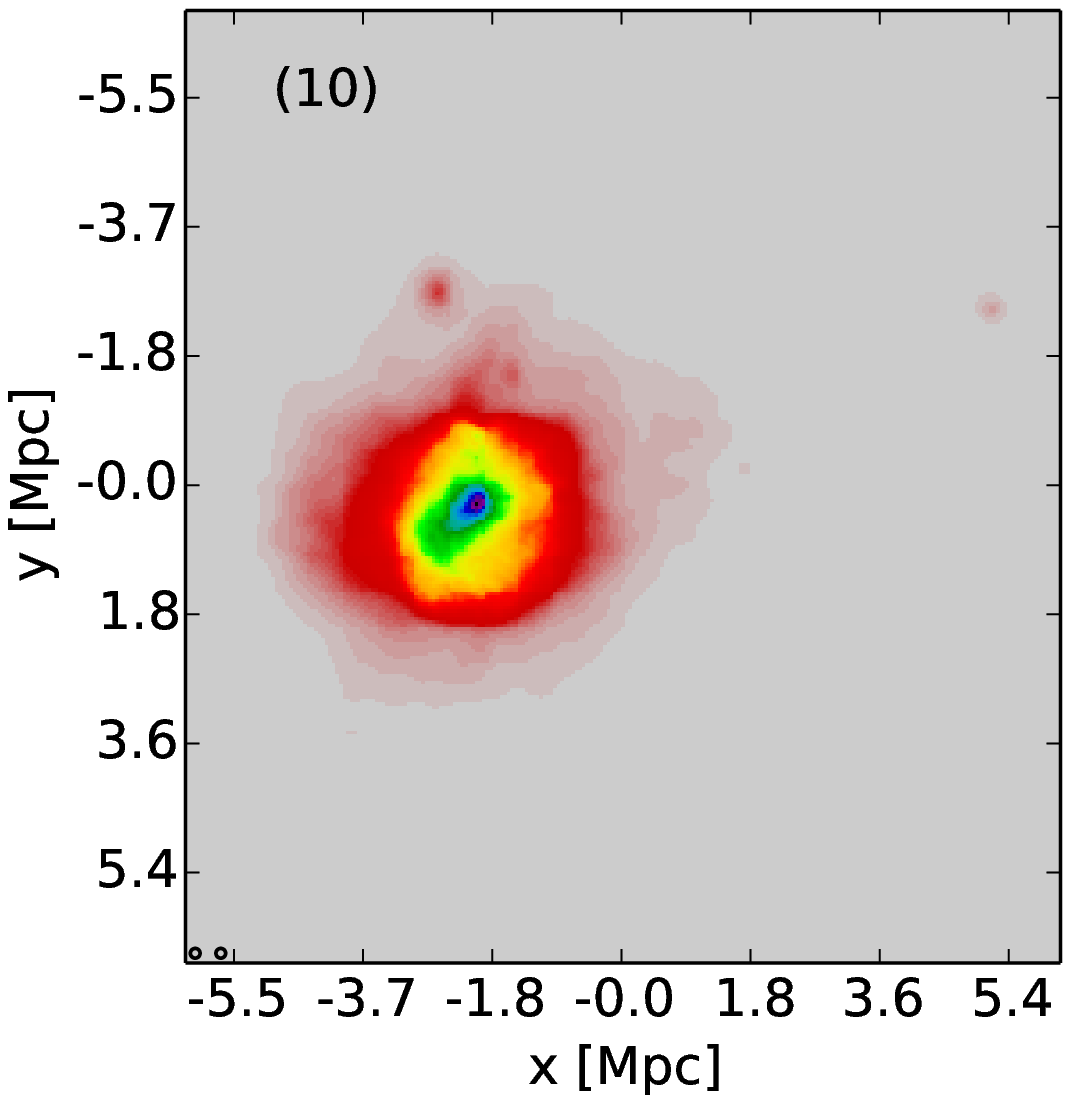}
\includegraphics[width=\haloPlotWidth\textwidth]{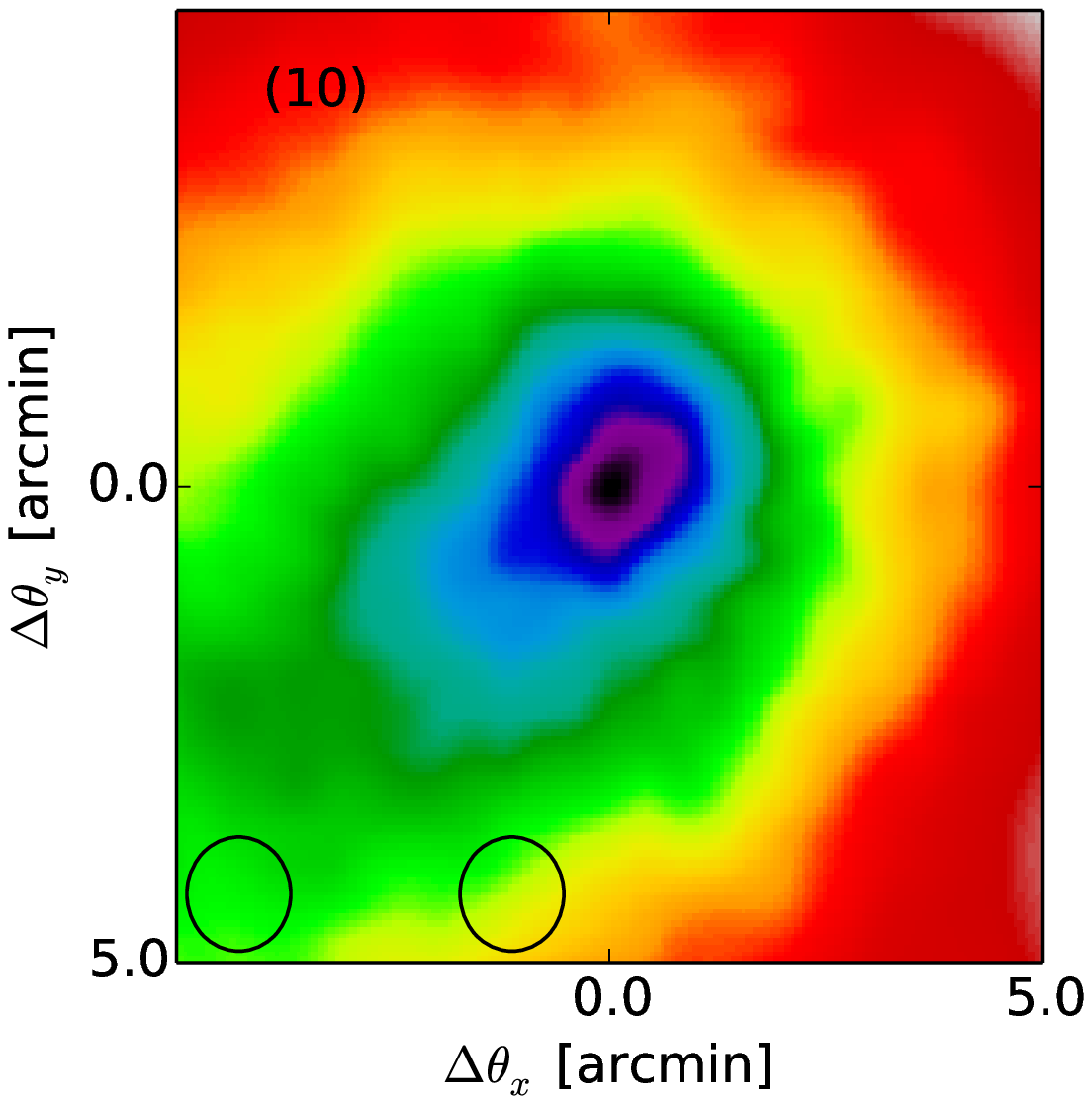}
\includegraphics[width=\haloPlotWidth\textwidth]{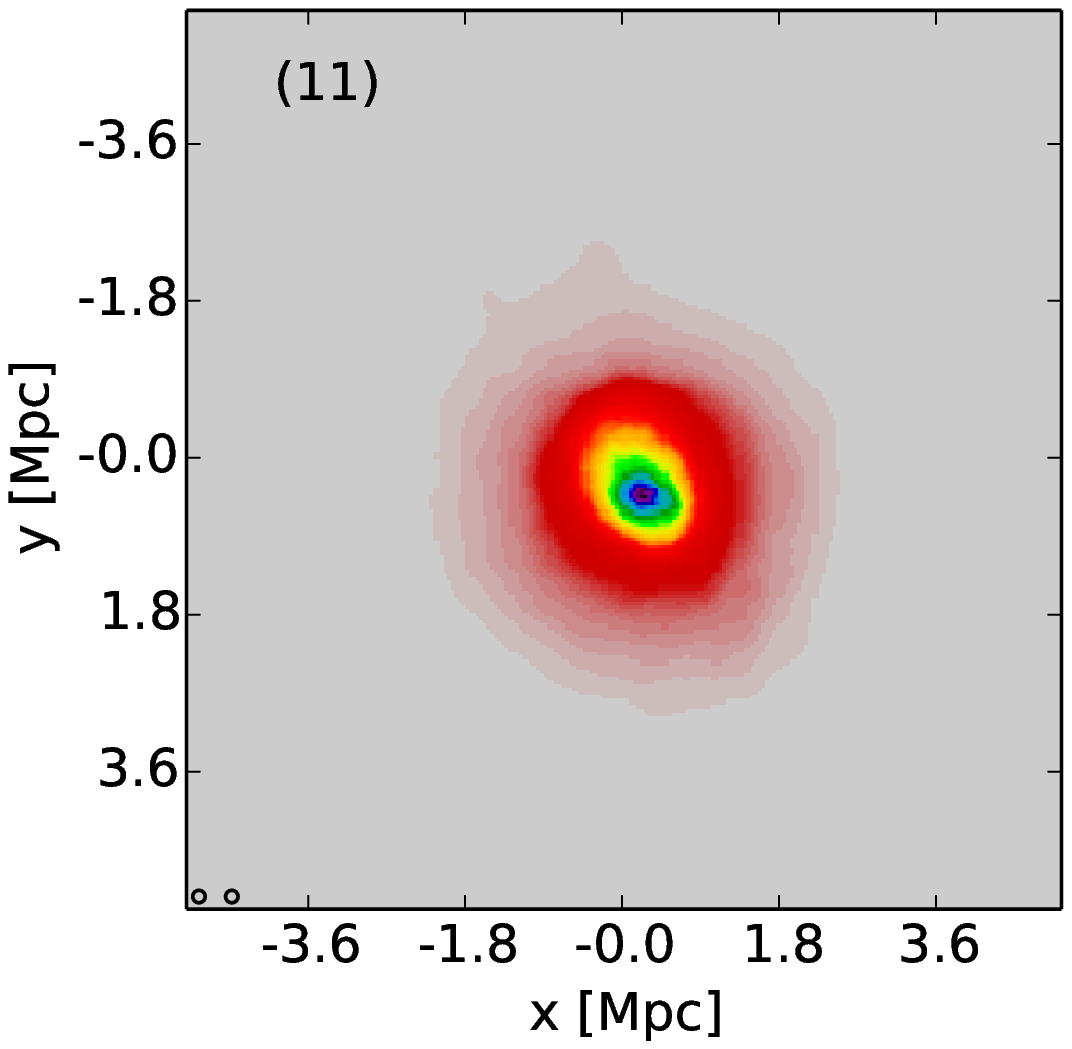}
\includegraphics[width=\haloPlotWidth\textwidth]{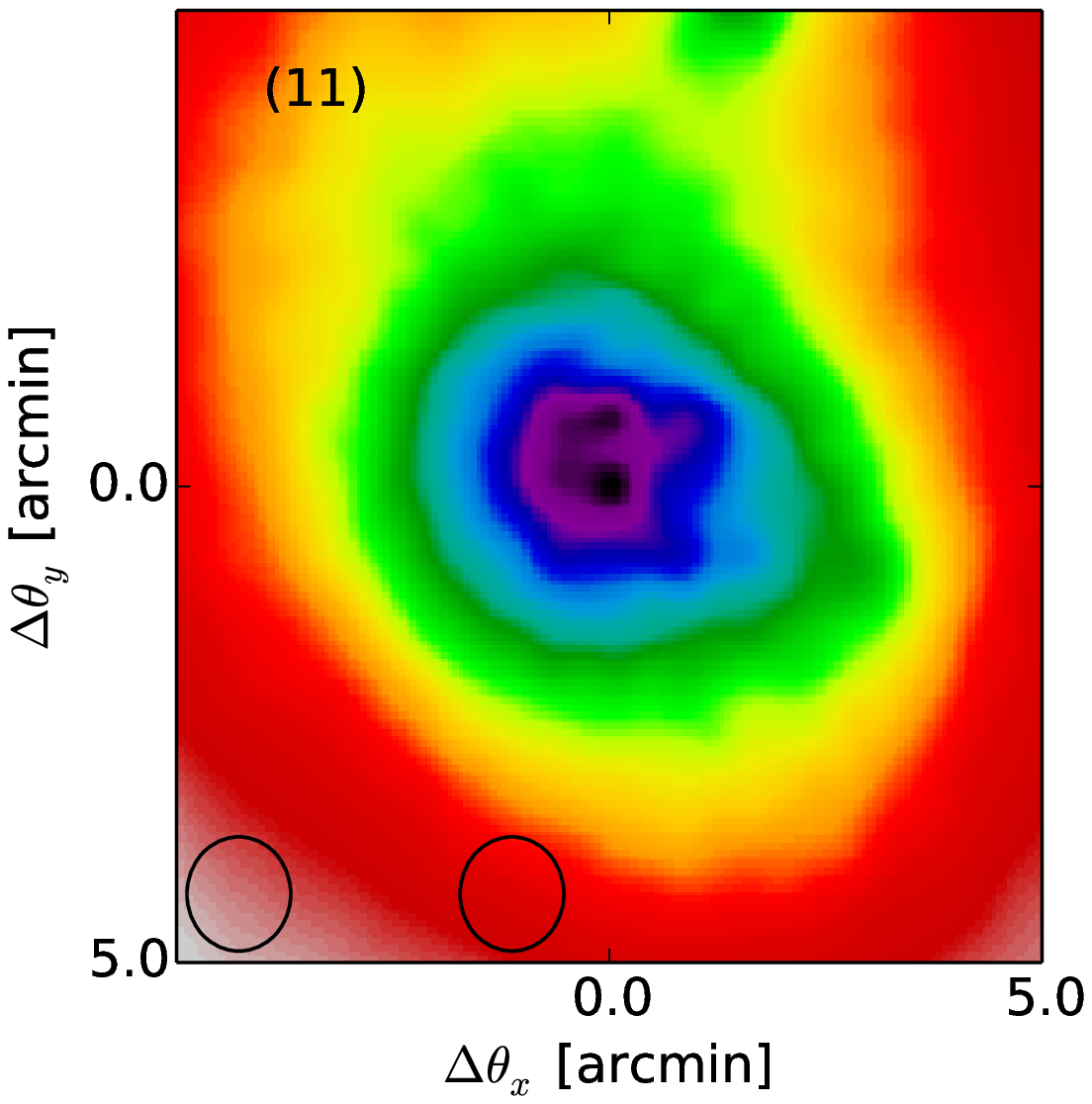}
\includegraphics[width=\haloPlotWidth\textwidth]{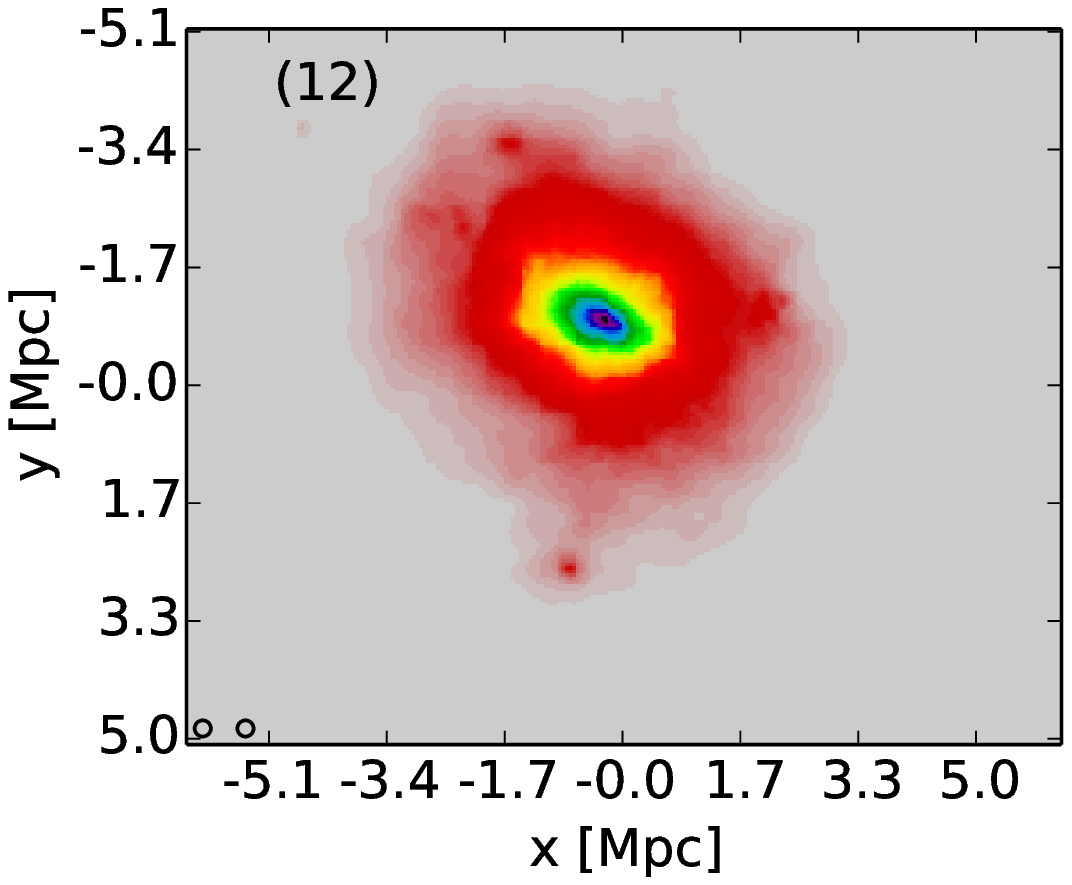}
\includegraphics[width=\haloPlotWidth\textwidth]{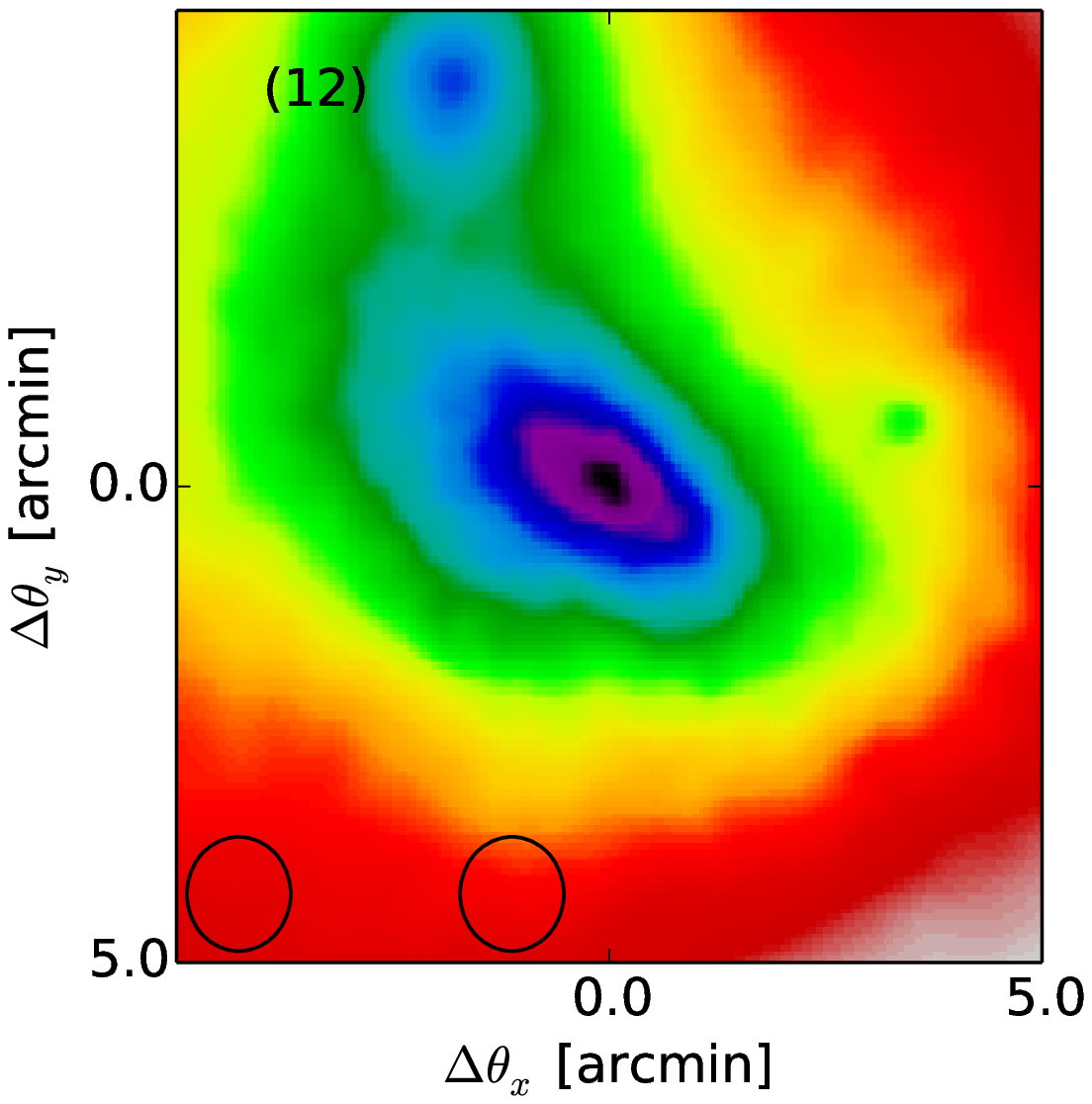}
\caption{Selection of simulated Compton $y$-parameter profiles
(in arbitrary units) for halos from Table.~\ref{tab:halos}. The
panels show profiles for individual halos in physical
coordinate space ({\em left}), and their coarse-grained version
obtained from high resolution maps in angular space with
contributions from other halos along the LOS ({\em right}).
The position of halos in the left-hand side panels is defined
by a box size that contains all FOF particles of the halo
associated with a given cluster. In the right-hand side panels
the SZ peak for the cluster is located in the plot center. For
any given cluster the flux density calculation is done at the
sky position of the peak. For each cluster the black circles
denote OCRA FWHMs and their relative separation ($s_{\rm
OCRA}$).  }
\label{fig:halos}
\end{figure}

We visually inspected all the clusters listed in
Table~\ref{tab:halos} and verified that each of the clusters from
rectangle ``1'' (halo IDs from 1 to 9, and 13) lie at sky
positions that are partially within another cluster's atmosphere
and also within the angular distance of the reference beam. An
example of such overlap is shown in Fig.~\ref{fig:halos} (top
panels).

Inspection of the three highest mass clusters in the \sampleHS
sample (halo IDs 10, 11, and 12; black dots in right panel of
Fig.~\ref{fig:f_z}) show that two of them (IDs 11 and 12) are
also affected to some degree by a LOS projection, but the
morphology of halo 10 shows no signs of another halo in the
composite high-resolution map.  Instead, the SZ signature has a
disturbed morphology with angular extents larger than a single
OCRA beam separation even though all three are at redshift
$z>0.1$. This results in small $F$ values, and motivates
measurements at larger angular separations.

\begin{figure}[!t]
\centering
\includegraphics[width=\haloPlotWidth\textwidth]{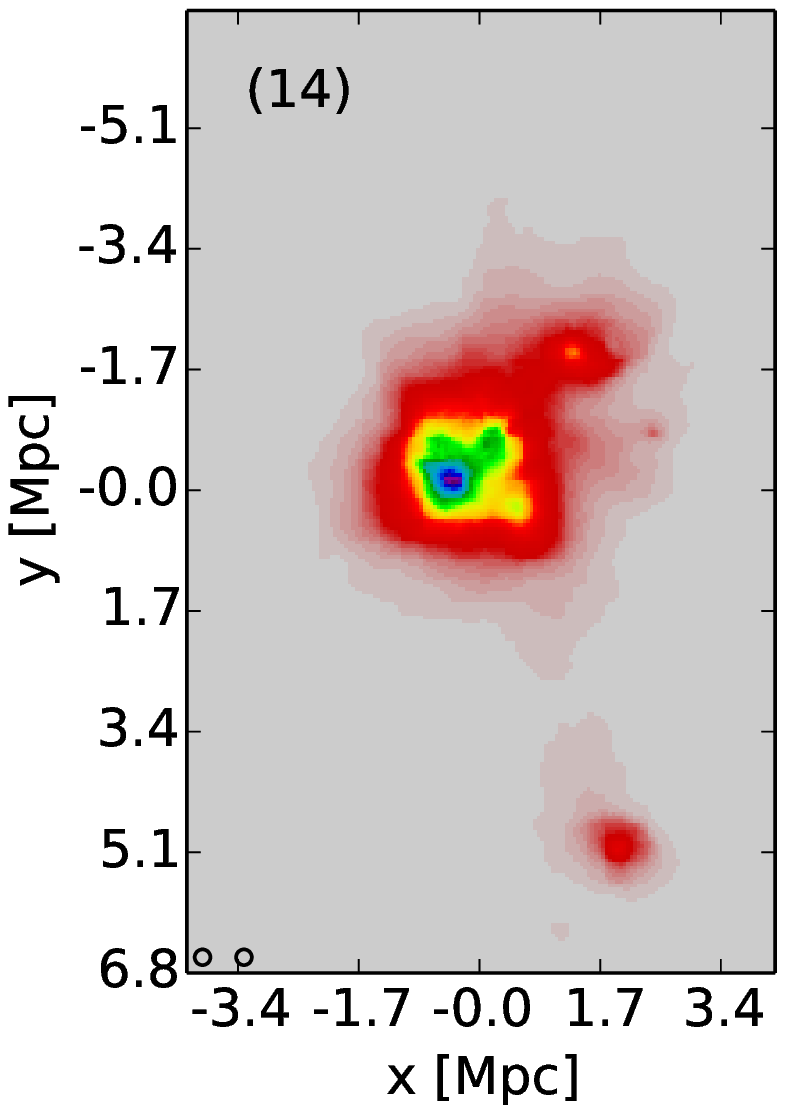}
\includegraphics[width=\haloPlotWidth\textwidth]{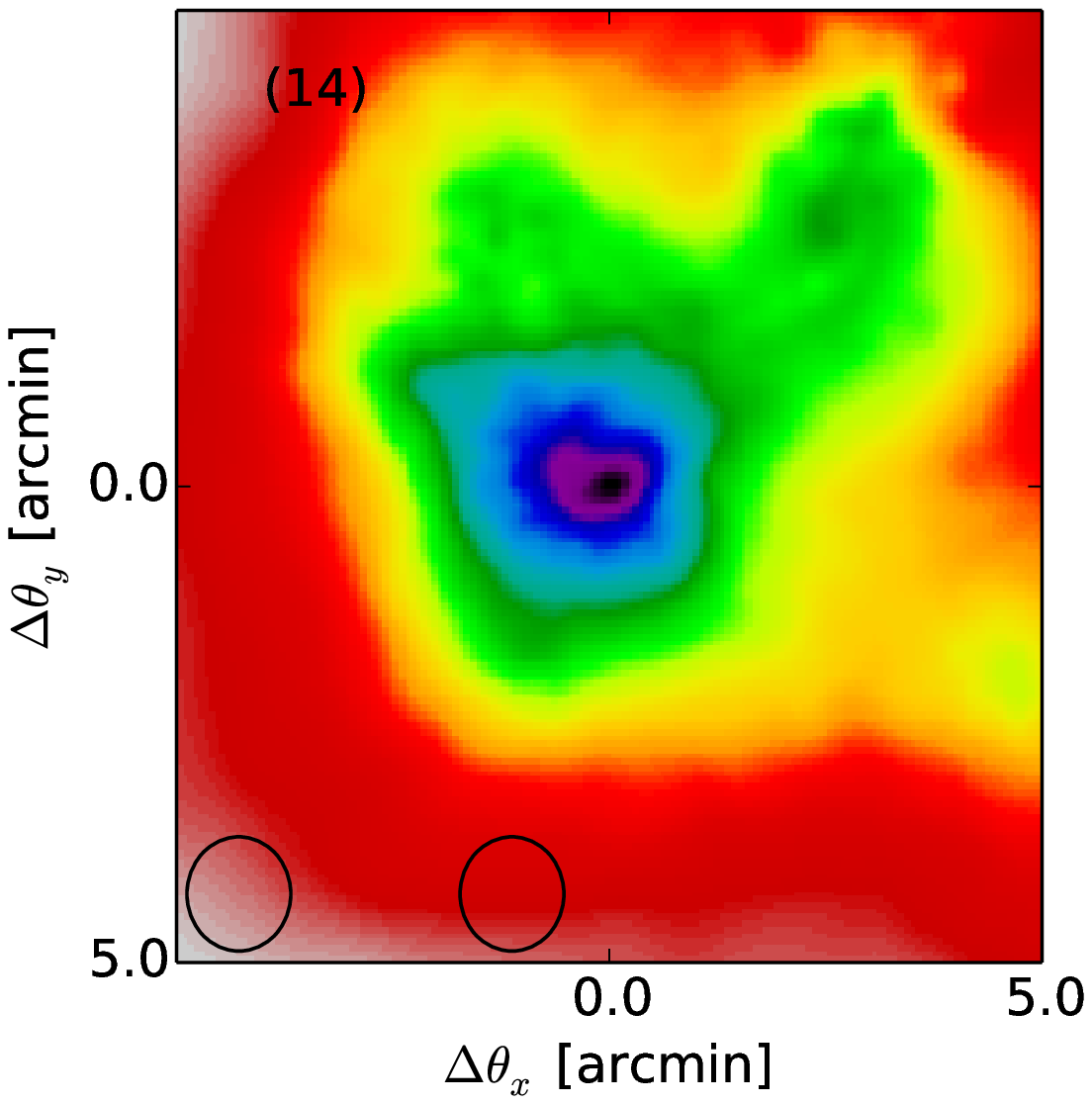}
\includegraphics[width=\haloPlotWidth\textwidth]{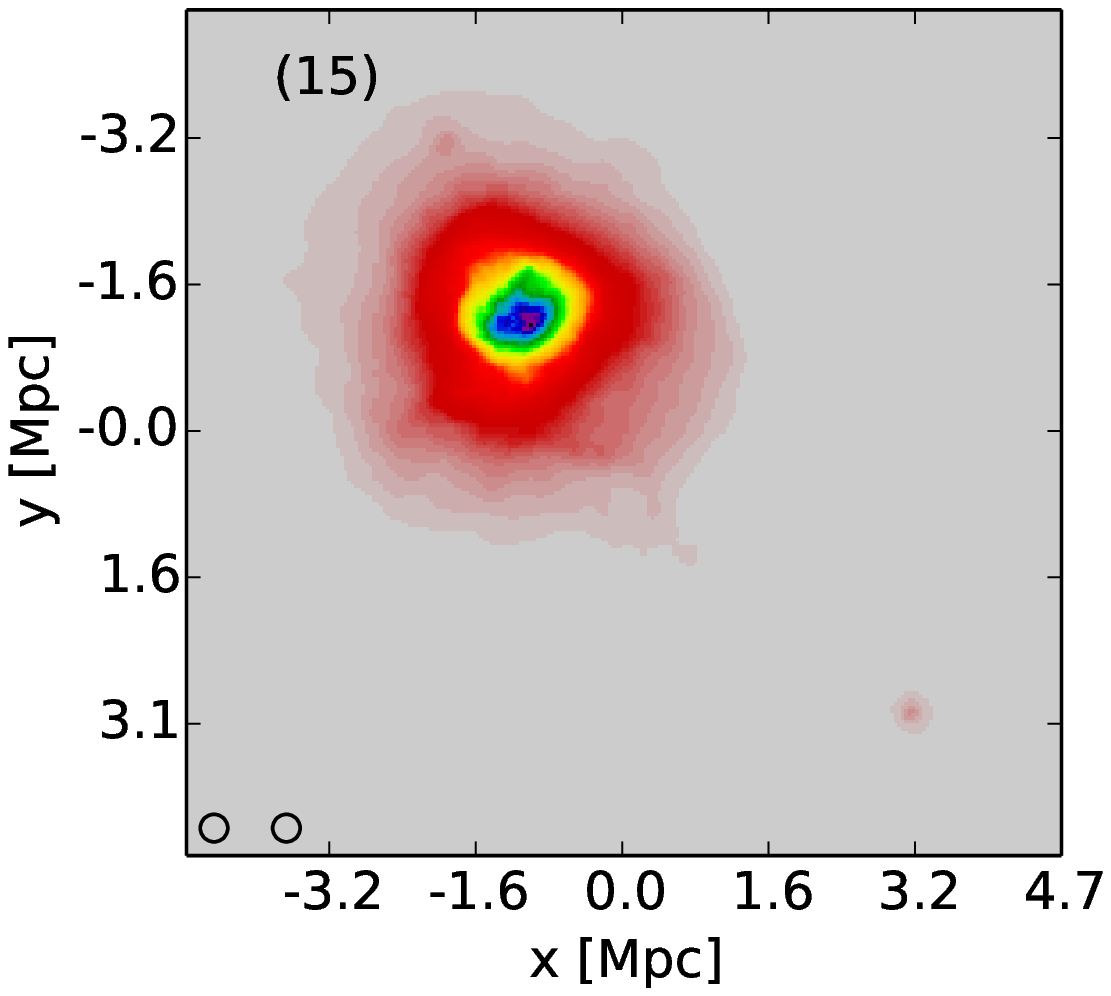}
\includegraphics[width=\haloPlotWidth\textwidth]{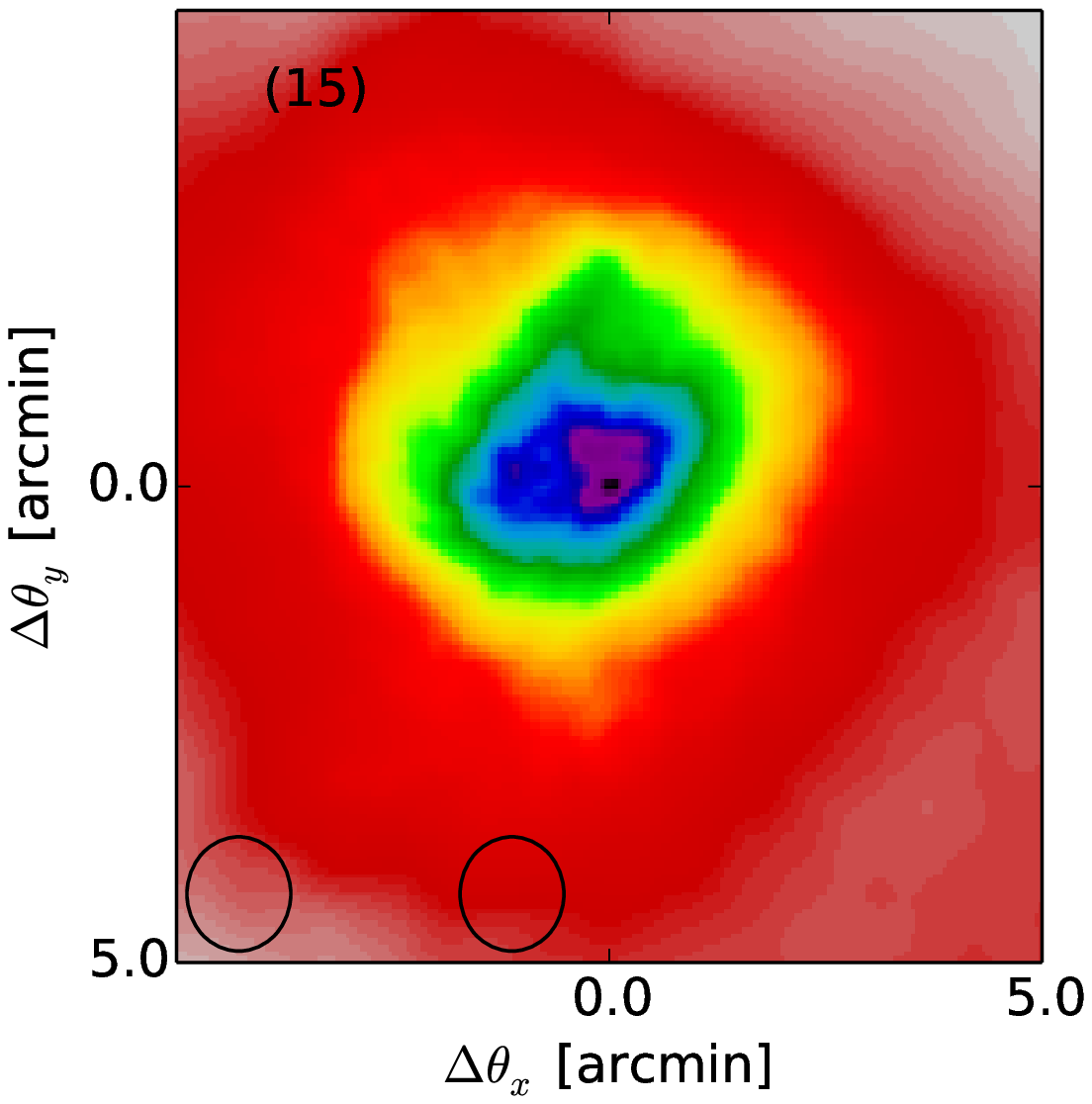}
\includegraphics[width=\haloPlotWidth\textwidth]{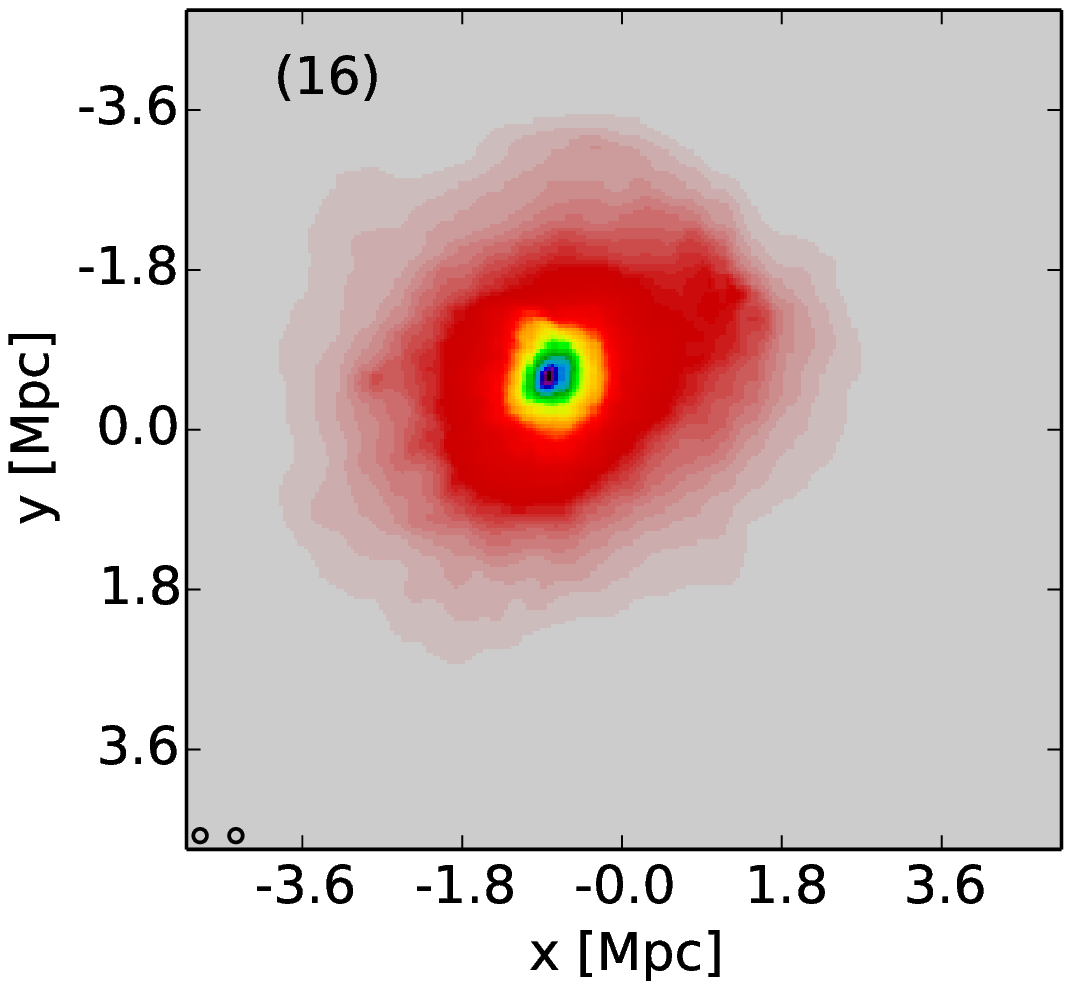}
\includegraphics[width=\haloPlotWidth\textwidth]{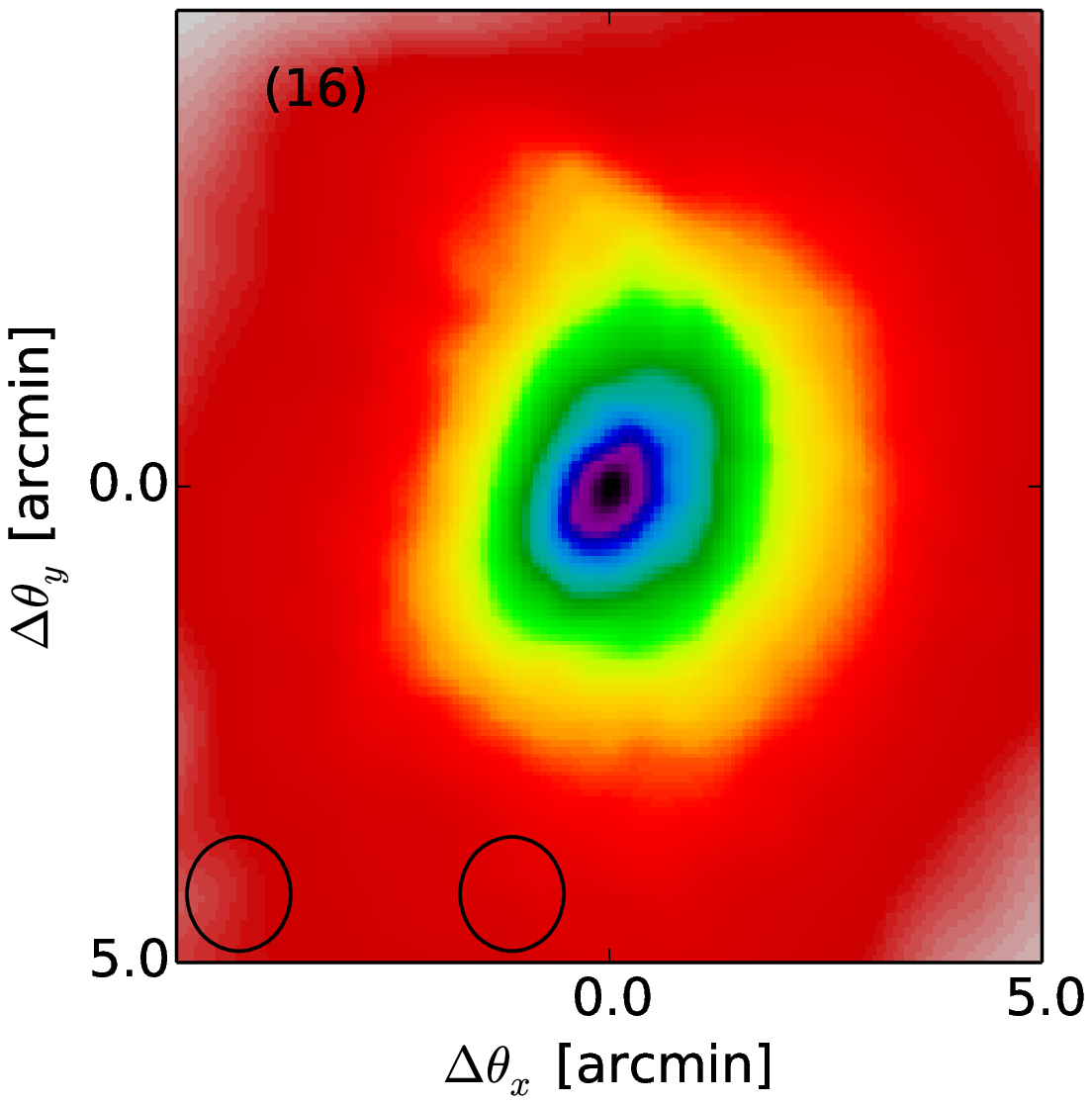}
\caption{As in Fig.~\ref{fig:halos} but for the selection of halos from rectangle ``2''. See Table.~\ref{tab:halos} for details.}
\label{fig:halos2}
\end{figure}

In order to test whether high redshift clusters that
significantly contribute to the scatter in the $F$--$z$ plane
(Fig.~\ref{fig:f_z}) could also benefit from observations out to
angular distances beyond $s_{\rm OCRA}$, we investigate the three
most massive clusters from rectangle ``2'' (Fig.~\ref{fig:f_z},
IDs: 14,15 and 16).  Their corresponding $F$ values
(Tab.~\ref{tab:halos}) do not seem to result from projection
effects. Instead, these clusters have extended atmospheres and/or
strongly disturbed and asymmetric SZ profiles (e.g. cluster 14,
Fig.~\ref{fig:halos2}).

Some of the heavy clusters have surface brightness profiles
(Fig.~\ref{fig:halos2}) that are strongly inconsistent with an
axially-symmetric $\beta$-profile. This necessitates using more
sophisticated two-dimensional profiles at the data analysis stage
\citep{Lancaster2011,Mirakhor2016}.  Clearly, heavy halos
generate low $F$ values and require large flux density
corrections with an OCRA type standard observational strategy
(Sec.~\ref{sec:strategy}).  These low $F$ values may partially
stem from spurious projection effects (e.g. halos 11 and 12)
which arise at the FOV generation stage for halos from the
\sampleHS sample (see Sec.~\ref{sec:discussion}).

The outlying halos (rectangle ``1'') are either mergers (close
pairs of SZ-strong halos), or have elongated of disturbed
morphology (e.g. halos 3 and 4), or have small scale
sub-structures.  However, in many cases these properties occur
at spatial scales that will not be resolved in OCRA~SZ
observations and/or may be relevant only as galaxy scale SZ
effects that are too faint to be detected.

\subsection{Practical aspects of using CMB templates}
\label{sec:template}

\begin{figure}[!t]
\centering
\includegraphics[width=0.47\textwidth]{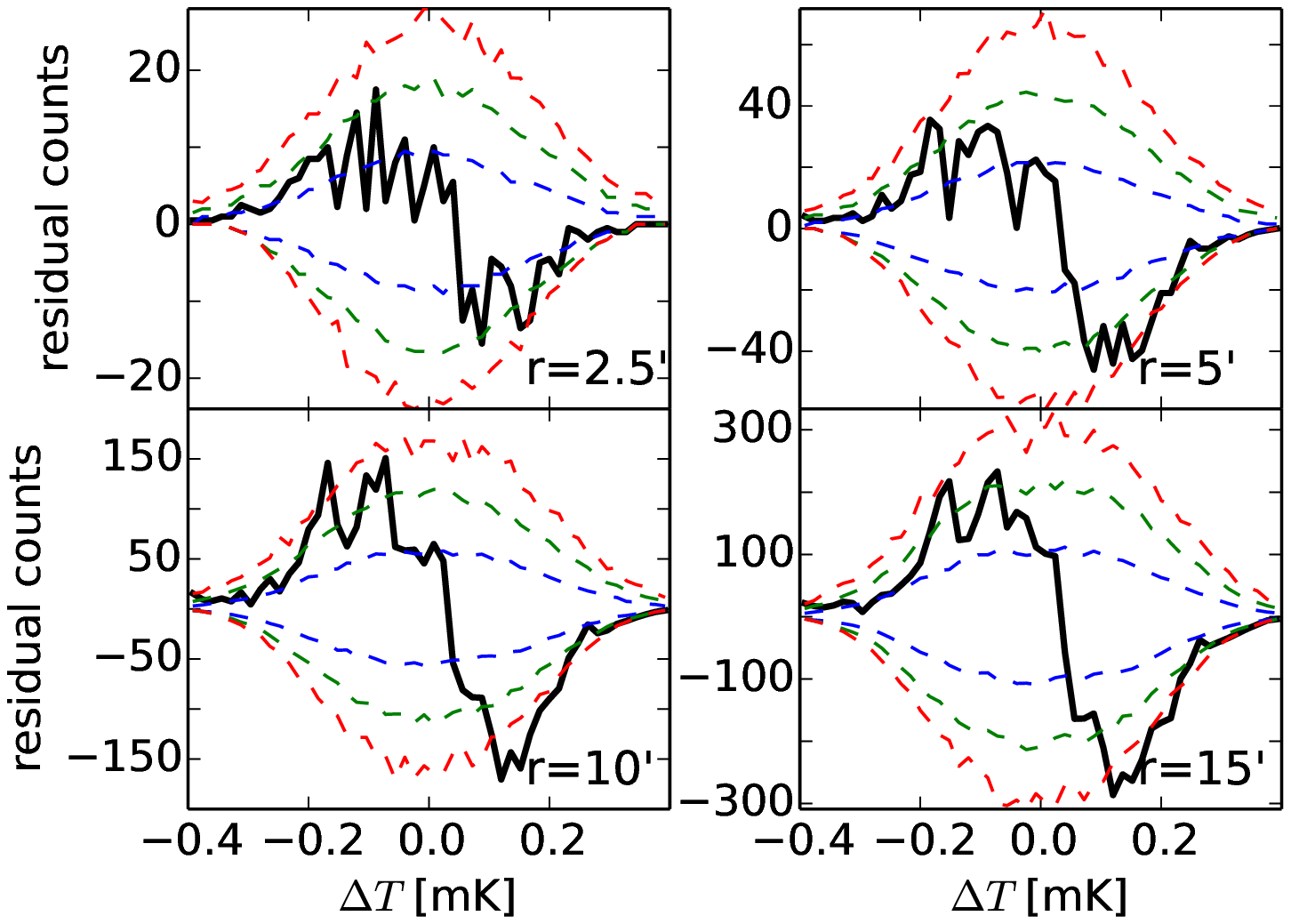}
\caption{Residual histogram (i.e. observational pixel
frequencies minus median pixel frequencies estimated from an
ensemble of Gaussian NILC map simulations) of the CMB
temperature fluctuations at and around the {\em Planck} SZ
galaxy clusters, measured in the {\em Planck} NILC inside
circular apertures of radius $r$ centered at the clusters'
positions (solid); and $1\sigma$, $2\sigma$ and $3\sigma$
confidence contours of the pixel frequencies in these
simulations (dashed).}
\label{fig:NILC-test}
\end{figure}

\begin{figure}[!t]
\centering
\includegraphics[width=0.47\textwidth]{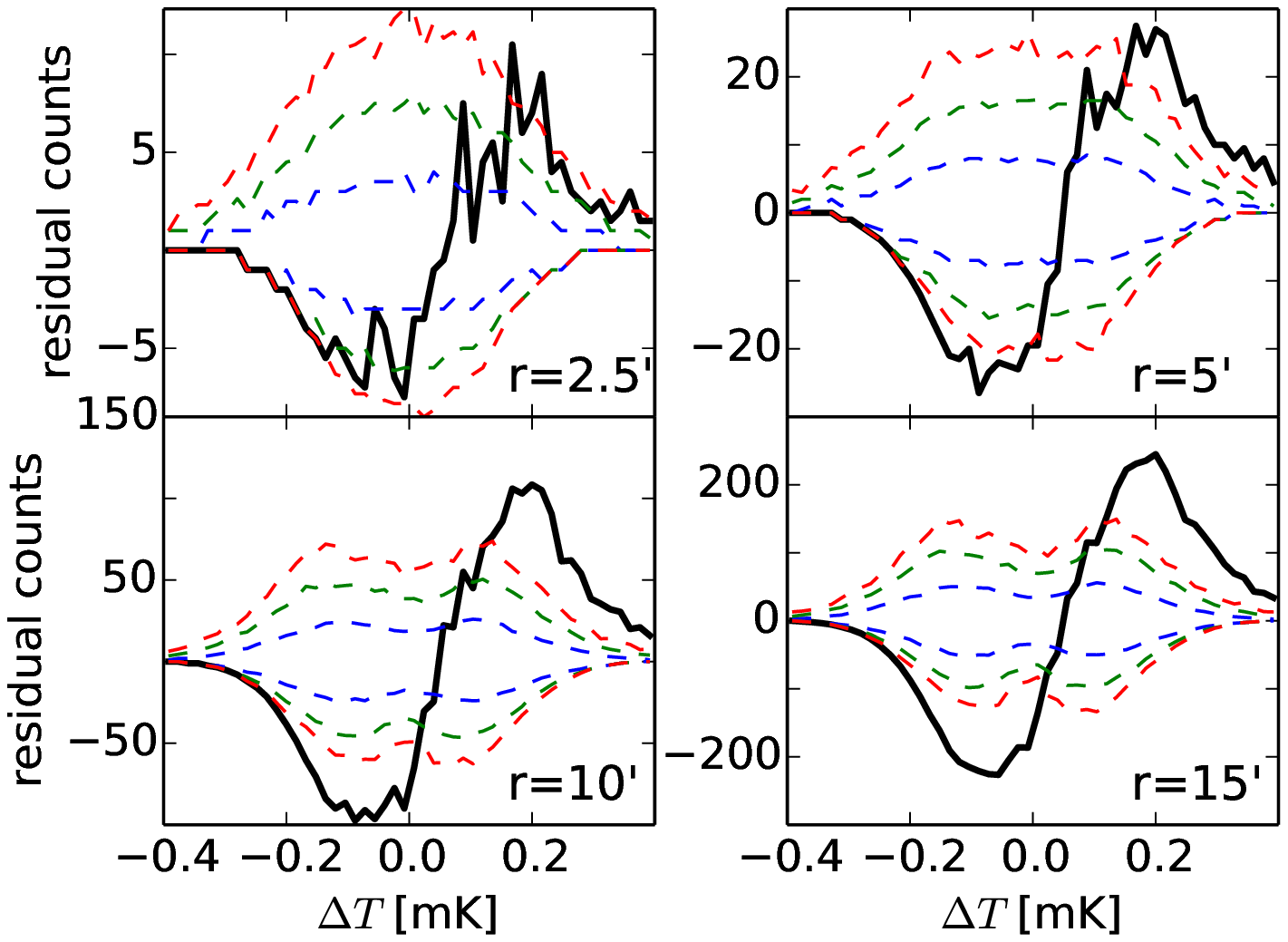}
\caption{As in Fig.~\ref{fig:NILC-test} but for the {\em Planck}
217~GHz frequency map and only for clusters at galactic
latitude $b>60^\circ$.}
\label{fig:217GHz-test}
\end{figure}

By subtracting the templated version of the CMB map
(Sec.~\ref{sec:CMBsims}) from the pure CMB simulation, it is easy
to estimate the upper limit of the residual CMB signal captured
in the OCRA difference beam observations.  The {\em Planck}
217~GHz and NILC maps have enough pixels to create a template of
resolution of the order of an arcminute, and at least the former
should contain only a negligible tSZ signal.

Although subtracting CMB templates from the pure CMB maps
decreases the large-scale variance by an order of magnitude, the
residual variance in the map is carried by high frequency noise
that will generate a small amount of dispersion in difference
observations (Fig.~\ref{fig:templates} bottom-right panel).
However, the residual small-scale noise should approximately
average out under rotation of the beams in the sky, and since the
large-scale power is effectively removed, increasing the
effective separation should not suffer from exponential variance
growth due to primordial CMB at arcminute angular scales.

How reliable are the 217~GHz or NILC {\em Planck} templates in
correcting single frequency SZ observations for confusion with
the primordial CMB?  The 217~GHz map is foreground contaminated
and the NILC map, although foreground cleaned, still may contain
residual tSZ signals at scales least optimized in the needlet
space.

In order to quantify the foregrounds and residual tSZ
contaminaiton in each map, we calculate histograms of the
temperature fluctuation distribution outside of a mask that
removes the full sky except for the directions towards {\em
Planck}-detected galaxy clusters from the PCSS SZ union R.2.08
catalog \citep{PlanckCollaboration2015f}.  Each non-masked region
is a circular patch of radius $a=\{2.5',5',10',15'\}$.
Foregrounds will generate strong positive skewness in the
temperature distribution, while the presence of residual tSZ in
the NILC map should manifest itself by either a positive or
negative skew depending on the frequency weights in the internal
linear combination.

While the results of the test for the {\em Planck} NILC map
(Fig.~\ref{fig:NILC-test}) do not give strong deviations from
Gaussian simulations, the 217~GHz map generally does.  The data
are inconsistent with Gaussian simulations even at high galactic
latitudes (Fig.~\ref{fig:217GHz-test}), although the significance
of the foregrounds seems to depend on the size of the circular
patch. This implies that the 217~GHz frequency map cannot readily
be used to mitigate the confusion due to CMB in OCRA observations
without further assumptions on the foregrounds' frequency
dependence.  However, it should be interesting to quantify the
significance of the arcminute scale Galactic foregrounds at
30~GHz at high and intermediate latitudes for OCRA difference
observations with small beam separations.  {\em Planck}-LFI data
might also help reduce these foregrounds, though we do not study
this here.

Since the foreground cleaned NILC map is statistically consistent
with Gaussian simulations
\citep{PlanckCollaboration2015a,PlanckCollaboration2015d} towards
the {\em Planck}-detected galaxy clusters
(Fig.~\ref{fig:NILC-test}), it should also be suitable for
mitigating CMB confusion in OCRA observations in directions
outside of the mask where clusters undetected by {\em Planck}
lie.

\subsection{Pointing requirements}
\label{sec:pointing}

\begin{figure*}[!t]
\centering
\includegraphics[width=0.45\textwidth]{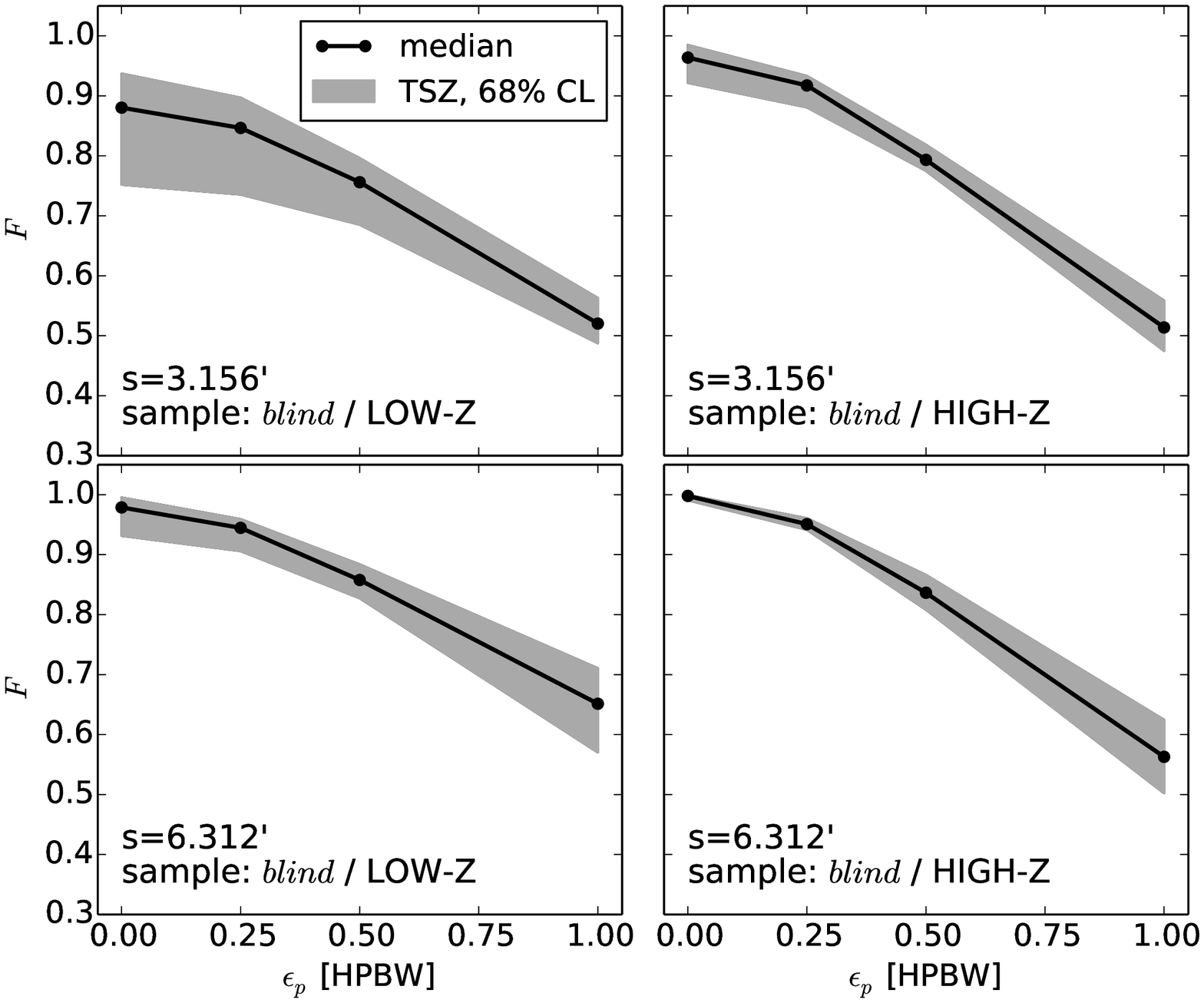}
\includegraphics[width=0.45\textwidth]{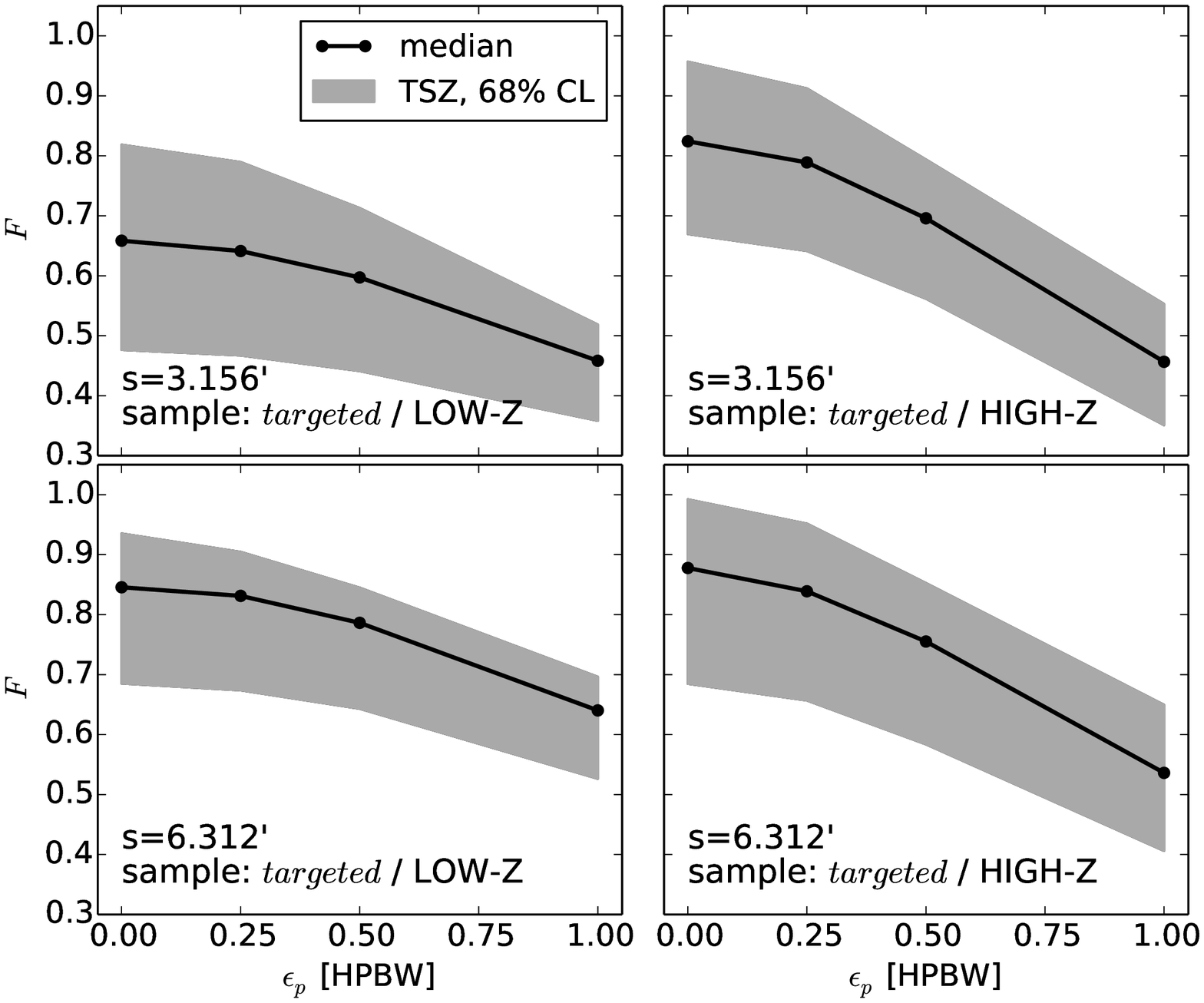}
\caption{Systematic effects in tSZ flux density reconstruction
from dual-beam observations as a function of telescope pointing
errors ($\epsilon_p$) and beam separation $s$ for the
\sampleDF~sample (left) and for the \sampleHS~sample (right).
The reference beam is assumed to cover all possible parallactic
angles for any given galaxy cluster.  }
\label{fig:F_vs_pointing}
\end{figure*}

In order to quantify the implications of telescope pointing
errors on the reconstruction of Compton $y$-parameters, and to
define pointing requirements, we introduce a pointing precision
parameter $\epsilon_p$ that defines the maximal angular distance
that a primary beam can have from the intended position, and then
we repeat the analysis of Sec.~\ref{sec:beam_separation}.  The
pointing error $p$ is drawn from a uniform distribution on
$[0,\epsilon_p]$, since the RT32 pointing and tracking are
dominated by systematic errors, and we investigate different
values of $\epsilon_p$.  Since galaxy cluster SZ profiles are
typically steep functions of angular separation, any pointing
inaccuracy will lead to biasing measurements of the central
comptonization parameter when taking averages from multiple
observational sequences.

Figure~\ref{fig:F_vs_pointing} shows that for the \sampleHS
sample, i.e. typically heavy clusters, pointing error up to
$\epsilon_p\approx {\rm HPBW}/4$ should not lead to strong
($>10\%$) extra biases relative to the $\epsilon_p=0$ case.
Measurements of the \sampleDF sample, i.e. typically less massive
clusters, are more sensitive to pointing errors, but if the
pointing accuracy is better than $\epsilon_p=0.005^\circ$ (${\rm
\theta_b^{\rm OCRA}}\approx 1.2'$) the additional systematic
effects will be smaller than $10\%$.  However, larger pointing
errors should be taken into account at the data analysis stage. An
observational campaign is currently under way to improve RT32 pointing
accuracy.

\section{Discussion}
\label{sec:discussion}

The map-making procedure that has been tested for recovering the
source intensity distribution from OCRA difference measurements
($s=s_{\rm OCRA}$) assumes a flat background. This is not a
problem for reconstructions of comptonization parameters from SZ
observations of heavy clusters, as with the standard OCRA
beam-pair separation the corrections due to CMB background are
small.  However, reconstructing cluster SZ profiles out to larger
angular distances could benefit from correcting the difference
measurements according to the {\em Planck} CMB template.  For
example, Fig.~\ref{fig:medianfHS} (left panel) shows that an
observation at $s=2 s_{\rm OCRA}$ decreases bias by $\Delta
F\approx 0.2$. At the same time CMB confusion broadens the 68\%
CR by $\Delta F \approx 0.1$, but applying a {\em Planck} CMB
template reverses this effect almost down to the intrinsic
tSZ+kSZ scatter.

As discussed in Sec.~\ref{sec:sample} the \sampleDF and \sampleHS
samples represent quite opposite observational
approaches. However, since halos of the \sampleHS sample were
selected from full simulation volumes (rather than from
light-cone sections), mock maps for this sample contain clusters
with angular sizes calculated according to their redshifts and
physical extents, as in the case of FOV simulations, but are
placed in the map at rectilinearly projected locations.  This
contaminates the resulting maps with halos that would not fall
into the assumed FOV in the standard light-cone approach.  These
spurious halo--halo overlaps may somewhat enlarge the 68\% CR
contours of various $F$ distributions (e.g. x=tSZ or x=tSZ+CMB).
A possible modification of the calculation scheme for the
\sampleHS sample would be to consider each halo independently,
thus completely ignoring the intrinsic projection effects that
exist in the light-cone approach, or by extending the FOV to a
hemisphere (which would probably require implementing adaptive
resolution maps to maintain the angular resolution of the present
calculations).

The cluster samples that we analyze were not screened to select
virialized clusters.  Although we analyzed sub-samples selected
using a virialization criterion (based on ratios of potential to
kinetic energy of FOF halo particles) the results presented here
are based on the full sample in order to retain a morphological
variety of SZ galaxy cluster profiles (Figs.~\ref{fig:halos} and
~\ref{fig:halos2}), and to expose the complexity of SZ flux
density reconstructions from observations that do not intend to
create multi-pixel intensity maps.

\section{Conclusions}
\label{sec:conclusions}

We quantify the significance of systematic effects arising in
dual-beam, differential observations of Sunyaev-Zel'dovich (SZ)
effect in galaxy clusters.  We primarily focus on effects
relevant to the reconstruction of comptonization parameters from
single frequency flux-density observations performed with the One
Centimeter Receiver Array (OCRA) -- a focal plane receiver with
arcminute scale beamwidths and arcminute scale beam separations
-- installed on the 32~m radio telescope in Toru\'n.

Using numerical simulations of large scale structure formation we
generate mock cluster samples (i) from blind surveys in small
fields of view and (ii) from volume limited targeted observations
of the most massive clusters (Sec.~\ref{sec:sample}).  Using mock
intensity maps of SZ effects we compare the true and recovered SZ
flux densities and quantify systematic effects caused by the
small beam separation, by primary CMB confusion and by
telescope pointing accuracy.

We find that for massive clusters the primary CMB confusion does
not significantly affect the recovered SZ effect flux density
with OCRA beam angular separation of $\approx 3'$. However, these
observations require large corrections due to the differential
observing strategy. On the other hand, measurements of SZ-faint
(or high redshift $0.4<z<1.0$) clusters may have their SZ
photometry erroneously estimated by $10\%$ or more due to CMB
confusion, which becomes stronger in observations that map larger
angular distances from cluster centers.

We investigate the possibility of mitigating the CMB confusion in
SZ observations that map scales beyond $3'$ from cluster centers
by using {\em Planck} CMB 217~GHz and foreground reduced NILC
maps as primary CMB templates. Using simulations we find that
these templates have sufficiently high angular resolution/low
noise to significantly mitigate CMB confusion in 30~GHz
observations of high-$z$ clusters, given that the templates do
not contain residual foregrounds. Using a simple one-point
statistic, targeted towards directions of known clusters, we
verify that at least the {\em Planck} NILC map should also be
sufficiently free from foregrounds to serve as a primary CMB
template (Sec.~\ref{sec:template}) that could improve OCRA-SZ or
similar observations extended to larger angular scales
(Sec.~\ref{sec:strategy}).

Finally, we find that RT32 telescope pointing and tracking
accuracy $\epsilon_p\ < 0.005^\circ$ should keep systematic
errors in recovered SZ flux densities (comptonization parameters)
below $\approx 5\%$ (after correcting for other systematic
effects) even in observations of SZ-weak clusters
(Sec.~\ref{sec:pointing}).

\section*{Acknowledgments}
Thank you to Mark Birkinshaw for discussion on OCRA observational
strategies, and to an anonymous referee for useful comments.
This research has made use of a modified version of the
GPL-licensed ccSHT library.  We also acknowledge use of the
matplotlib plotting library \citep{Hunter2007}.
This work was financially supported by the Polish National
Science Centre through grant DEC-2011/03/D/ST9/03373.  A part of
this project has made use of computations made under grant 197 of
the Pozna{\'n} Supercomputing and Networking Center (PSNC).

\bibliography{bibliography} 

\begin{thebibliography}{35}
\expandafter\ifx\csname natexlab\endcsname\relax\def\natexlab#1{#1}\fi

\bibitem[{{Adam} {et~al.}(2014){Adam}, {Comis}, {Mac{\'{\i}}as-P{\'e}rez},
  {Adane}, {Ade}, {Andr{\'e}}, {Beelen}, {Belier}, {Beno{\^i}t}, {Bideaud},
  {Billot}, {Boudou}, {Bourrion}, {Calvo}, {Catalano}, {Coiffard}, {D'Addabbo},
  {D{\'e}sert}, {Doyle}, {Goupy}, {Kramer}, {Leclercq}, {Martino}, {Mauskopf},
  {Mayet}, {Monfardini}, {Pajot}, {Pascale}, {Perotto}, {Pointecouteau},
  {Ponthieu}, {Rev{\'e}ret}, {Rodriguez}, {Savini}, {Schuster}, {Sievers},
  {Tucker}, \& {Zylka}}]{Adam2014}
{Adam}, R., {Comis}, B., {Mac{\'{\i}}as-P{\'e}rez}, J.~F., {et~al.} 2014, \aap,
  569, A66, \eprint{1310.6237}

\bibitem[{{Bender} {et~al.}(2016){Bender}, {Kennedy}, {Ade}, {Basu},
  {Bertoldi}, {Burkutean}, {Clarke}, {Dahlin}, {Dobbs}, {Ferrusca}, {Flanigan},
  {Halverson}, {Holzapfel}, {Horellou}, {Johnson}, {Kermish}, {Klein},
  {Kneissl}, {Lanting}, {Lee}, {Mehl}, {Menten}, {Muders}, {Nagarajan},
  {Pacaud}, {Reichardt}, {Richards}, {Schaaf}, {Schwan}, {Sommer}, {Spieler},
  {Tucker}, \& {Westbrook}}]{Bender2016}
{Bender}, A.~N., {Kennedy}, J., {Ade}, P.~A.~R., {et~al.} 2016, \mnras, 460,
  3432

\bibitem[{{Birkinshaw} \& {Lancaster}(2005)}]{Birkinshaw2005}
{Birkinshaw}, M. \& {Lancaster}, K. 2005, in Background Microwave Radiation and
  Intracluster Cosmology, ed. F.~{Melchiorri} \& Y.~{Rephaeli}, Vol.
  2005937974, 127

\bibitem[{{Bleem} {et~al.}(2015){Bleem}, {Stalder}, {de Haan}, {Aird}, {Allen},
  {Applegate}, {Ashby}, {Bautz}, {Bayliss}, {Benson}, {Bocquet}, {Brodwin},
  {Carlstrom}, {Chang}, {Chiu}, {Cho}, {Clocchiatti}, {Crawford}, {Crites},
  {Desai}, {Dietrich}, {Dobbs}, {Foley}, {Forman}, {George}, {Gladders},
  {Gonzalez}, {Halverson}, {Hennig}, {Hoekstra}, {Holder}, {Holzapfel},
  {Hrubes}, {Jones}, {Keisler}, {Knox}, {Lee}, {Leitch}, {Liu}, {Lueker},
  {Luong-Van}, {Mantz}, {Marrone}, {McDonald}, {McMahon}, {Meyer}, {Mocanu},
  {Mohr}, {Murray}, {Padin}, {Pryke}, {Reichardt}, {Rest}, {Ruel}, {Ruhl},
  {Saliwanchik}, {Saro}, {Sayre}, {Schaffer}, {Schrabback}, {Shirokoff},
  {Song}, {Spieler}, {Stanford}, {Staniszewski}, {Stark}, {Story}, {Stubbs},
  {Vanderlinde}, {Vieira}, {Vikhlinin}, {Williamson}, {Zahn}, \&
  {Zenteno}}]{Bleem2015}
{Bleem}, L.~E., {Stalder}, B., {de Haan}, T., {et~al.} 2015, \apjs, 216, 27,
  \eprint{1409.0850}

\bibitem[{{Browne} {et~al.}(2000){Browne}, {Mao}, {Wilkinson}, {Kus},
  {Marecki}, \& {Birkinshaw}}]{Browne2000}
{Browne}, I.~W., {Mao}, S., {Wilkinson}, P.~N., {et~al.} 2000, in Society of
  Photo-Optical Instrumentation Engineers (SPIE) Conference Series, Vol. 4015,
  Radio Telescopes, ed. H.~R. {Butcher}, 299--307

\bibitem[{{Calvo} {et~al.}(2016){Calvo}, {Beno{\^i}t}, {Catalano}, {Goupy},
  {Monfardini}, {Ponthieu}, {Barria}, {Bres}, {Grollier}, {Garde}, {Leggeri},
  {Pont}, {Triqueneaux}, {Adam}, {Bourrion}, {Mac{\'{\i}}as-P{\'e}rez},
  {Rebolo}, {Ritacco}, {Scordilis}, {Tourres}, {Adane}, {Coiffard}, {Leclercq},
  {D{\'e}sert}, {Doyle}, {Mauskopf}, {Tucker}, {Ade}, {Andr{\'e}}, {Beelen},
  {Belier}, {Bideaud}, {Billot}, {Comis}, {D'Addabbo}, {Kramer}, {Martino},
  {Mayet}, {Pajot}, {Pascale}, {Perotto}, {Rev{\'e}ret}, {Ritacco},
  {Rodriguez}, {Savini}, {Schuster}, {Sievers}, \& {Zylka}}]{Calvo2016}
{Calvo}, M., {Beno{\^i}t}, A., {Catalano}, A., {et~al.} 2016, Journal of Low
  Temperature Physics, 184, 816, \eprint{1601.02774}

\bibitem[{{Dawson} {et~al.}(2006){Dawson}, {Holzapfel}, {Carlstrom}, {Joy}, \&
  {LaRoque}}]{Dawson2006}
{Dawson}, K.~S., {Holzapfel}, W.~L., {Carlstrom}, J.~E., {Joy}, M., \&
  {LaRoque}, S.~J. 2006, \apj, 647, 13, \eprint{astro-ph/0602413}

\bibitem[{{Delabrouille} {et~al.}(2009){Delabrouille}, {Cardoso}, {Le Jeune},
  {Betoule}, {Fay}, \& {Guilloux}}]{Delabrouille2009}
{Delabrouille}, J., {Cardoso}, J.-F., {Le Jeune}, M., {et~al.} 2009, \aap, 493,
  835, \eprint{0807.0773}

\bibitem[{{Dobbs} {et~al.}(2006){Dobbs}, {Halverson}, {Ade}, {Basu}, {Beelen},
  {Bertoldi}, {Cohalan}, {Cho}, {G{\"u}sten}, {Holzapfel}, {Kermish},
  {Kneissl}, {Kov{\'a}cs}, {Kreysa}, {Lanting}, {Lee}, {Lueker}, {Mehl},
  {Menten}, {Muders}, {Nord}, {Plagge}, {Richards}, {Schilke}, {Schwan},
  {Spieler}, {Weiss}, \& {White}}]{Dobbs2006}
{Dobbs}, M., {Halverson}, N.~W., {Ade}, P.~A.~R., {et~al.} 2006, \nar, 50, 960

\bibitem[{{G{\'o}rski} {et~al.}(2005){G{\'o}rski}, {Hivon}, {Banday},
  {Wandelt}, {Hansen}, {Reinecke}, \& {Bartelmann}}]{Gorski2005}
{G{\'o}rski}, K.~M., {Hivon}, E., {Banday}, A.~J., {et~al.} 2005, \apj, 622,
  759, \eprint{astro-ph/0409513}

\bibitem[{{Hasselfield} {et~al.}(2013){Hasselfield}, {Hilton}, {Marriage},
  {Addison}, {Barrientos}, {Battaglia}, {Battistelli}, {Bond}, {Crichton},
  {Das}, {Devlin}, {Dicker}, {Dunkley}, {D{\"u}nner}, {Fowler}, {Gralla},
  {Hajian}, {Halpern}, {Hincks}, {Hlozek}, {Hughes}, {Infante}, {Irwin},
  {Kosowsky}, {Marsden}, {Menanteau}, {Moodley}, {Niemack}, {Nolta}, {Page},
  {Partridge}, {Reese}, {Schmitt}, {Sehgal}, {Sherwin}, {Sievers}, {Sif{\'o}n},
  {Spergel}, {Staggs}, {Swetz}, {Switzer}, {Thornton}, {Trac}, \&
  {Wollack}}]{Hasselfield2013}
{Hasselfield}, M., {Hilton}, M., {Marriage}, T.~A., {et~al.} 2013, \jcap, 7, 8,
  \eprint{1301.0816}

\bibitem[{Hunter(2007)}]{Hunter2007}
Hunter, J.~D. 2007, Computing In Science \& Engineering, 9, 90

\bibitem[{{Lancaster} {et~al.}(2011){Lancaster}, {Birkinshaw}, {Gawro{\'n}ski},
  {Battye}, {Browne}, {Davis}, {Giles}, {Feiler}, {Kus}, {Lew}, {Lowe},
  {Maughan}, {Alareedh}, {Pazderska}, {Pazderski}, {Peel}, {Roukema}, \&
  {Wilkinson}}]{Lancaster2011}
{Lancaster}, K., {Birkinshaw}, M., {Gawro{\'n}ski}, M.~P., {et~al.} 2011,
  \mnras, 418, 1441, \eprint{1106.3766}

\bibitem[{{Lancaster} {et~al.}(2007){Lancaster}, {Birkinshaw}, {Gawro{\'n}ski},
  {Browne}, {Feiler}, {Kus}, {Lowe}, {Pazderski}, \&
  {Wilkinson}}]{Lancaster2007}
{Lancaster}, K., {Birkinshaw}, M., {Gawro{\'n}ski}, M.~P., {et~al.} 2007,
  \mnras, 378, 673, \eprint{0705.3336}

\bibitem[{{Lew} {et~al.}(2015){Lew}, {Birkinshaw}, {Wilkinson}, \&
  {Kus}}]{Lew2015}
{Lew}, B., {Birkinshaw}, M., {Wilkinson}, P., \& {Kus}, A. 2015, \jcap, 2, 4,
  \eprint{1410.3660}

\bibitem[{{Lew} \& {Uscka-Kowalkowska}(2016)}]{Lew2016}
{Lew}, B. \& {Uscka-Kowalkowska}, J. 2016, \mnras, 455, 2901,
  \eprint{1506.00225}

\bibitem[{{Lewis} {et~al.}(2000){Lewis}, {Challinor}, \& {Lasenby}}]{Lewis2000}
{Lewis}, A., {Challinor}, A., \& {Lasenby}, A. 2000, \apj, 538, 473,
  \eprint{astro-ph/9911177}

\bibitem[{{Lin} {et~al.}(2016){Lin}, {Nishioka}, {Wang}, {Locutus Huang},
  {Liao}, {Proty Wu}, {Koch}, {Umetsu}, {Chen}, {Chan}, {Chang}, {Lucky Chang},
  {Cheng}, {Duy}, {Fu}, {Han}, {Ho}, {Ho}, {Ho}, {Huang}, {Jiang}, {Kubo},
  {Li}, {Lin}, {Liu}, {Martin-Cocher}, {Molnar}, {Nunez}, {Oshiro}, {Pai},
  {Raffin}, {Ridenour}, {Shih}, {Stoebner}, {Teo}, {Yeh}, {Williams}, \&
  {Birkinshaw}}]{Lin2016}
{Lin}, K.-Y., {Nishioka}, H., {Wang}, F.-C., {et~al.} 2016, ArXiv e-prints,
  \eprint{1605.09261}

\bibitem[{{Mantz} {et~al.}(2014){Mantz}, {Abdulla}, {Carlstrom}, {Greer},
  {Leitch}, {Marrone}, {Muchovej}, {Adami}, {Birkinshaw}, {Bremer}, {Clerc},
  {Giles}, {Horellou}, {Maughan}, {Pacaud}, {Pierre}, \& {Willis}}]{Mantz2014}
{Mantz}, A.~B., {Abdulla}, Z., {Carlstrom}, J.~E., {et~al.} 2014, \apj, 794,
  157, \eprint{1401.2087}

\bibitem[{{Melin} {et~al.}(2006){Melin}, {Bartlett}, \&
  {Delabrouille}}]{Melin2006}
{Melin}, J.-B., {Bartlett}, J.~G., \& {Delabrouille}, J. 2006, \aap, 459, 341,
  \eprint{astro-ph/0602424}

\bibitem[{{Mirakhor} \& {Birkinshaw}(2016)}]{Mirakhor2016}
{Mirakhor}, M.~S. \& {Birkinshaw}, M. 2016, \mnras, 457, 2918,
  \eprint{1601.05304}

\bibitem[{{Muchovej} {et~al.}(2012){Muchovej}, {Leitch}, {Culverhouse},
  {Carpenter}, \& {Sievers}}]{Muchovej2012}
{Muchovej}, S., {Leitch}, E., {Culverhouse}, T., {Carpenter}, J., \& {Sievers},
  J. 2012, \apj, 749, 46, \eprint{1202.0527}

\bibitem[{{Peel}(2010)}]{Peel2010}
{Peel}, M. 2010, PhD thesis, ArXiv e-prints, \eprint{1006.2760}

\bibitem[{{Peel} {et~al.}(2011){Peel}, {Gawro{\'n}ski}, {Battye}, {Birkinshaw},
  {Browne}, {Davis}, {Feiler}, {Kus}, {Lancaster}, {Lowe}, {Pazderska},
  {Pazderski}, {Roukema}, \& {Wilkinson}}]{Peel2011}
{Peel}, M.~W., {Gawro{\'n}ski}, M.~P., {Battye}, R.~A., {et~al.} 2011, \mnras,
  410, 2690, \eprint{1007.5242}

\bibitem[{{Planck Collaboration} {et~al.}(2016{\natexlab{a}}){Planck
  Collaboration}, {Adam}, {Ade}, {Aghanim}, {Arnaud}, {Ashdown}, {Aumont},
  {Baccigalupi}, {Banday}, {Barreiro}, \& et~al.}]{PlanckCollaboration2016a}
{Planck Collaboration}, {Adam}, R., {Ade}, P.~A.~R., {et~al.}
  2016{\natexlab{a}}, \aap, 594, A9, \eprint{1502.05956}

\bibitem[{{Planck Collaboration} {et~al.}(2016{\natexlab{b}}){Planck
  Collaboration}, {Adam}, {Ade}, {Aghanim}, {Arnaud}, {Ashdown}, {Aumont},
  {Baccigalupi}, {Banday}, {Barreiro}, \& et~al.}]{PlanckCollaboration2016b}
{Planck Collaboration}, {Adam}, R., {Ade}, P.~A.~R., {et~al.}
  2016{\natexlab{b}}, \aap, 594, A8, \eprint{1502.01587}

\bibitem[{{Planck Collaboration} {et~al.}(2016{\natexlab{c}}){Planck
  Collaboration}, {Ade}, {Aghanim}, {Akrami}, {Aluri}, {Arnaud}, {Ashdown},
  {Aumont}, {Baccigalupi}, {Banday}, \& et~al.}]{PlanckCollaboration2015a}
{Planck Collaboration}, {Ade}, P.~A.~R., {Aghanim}, N., {et~al.}
  2016{\natexlab{c}}, \aap, 594, A16, \eprint{1506.07135}

\bibitem[{{Planck Collaboration} {et~al.}(2016{\natexlab{d}}){Planck
  Collaboration}, {Ade}, {Aghanim}, {Arg{\"u}eso}, {Arnaud}, {Ashdown},
  {Aumont}, {Baccigalupi}, {Banday}, {Barreiro}, \&
  et~al.}]{PlanckCollaboration2015f}
{Planck Collaboration}, {Ade}, P.~A.~R., {Aghanim}, N., {et~al.}
  2016{\natexlab{d}}, \aap, 594, A26, \eprint{1507.02058}

\bibitem[{{Planck Collaboration} {et~al.}(2016{\natexlab{e}}){Planck
  Collaboration}, {Ade}, {Aghanim}, {Arnaud}, {Arroja}, {Ashdown}, {Aumont},
  {Baccigalupi}, {Ballardini}, {Banday}, \& et~al.}]{PlanckCollaboration2015d}
{Planck Collaboration}, {Ade}, P.~A.~R., {Aghanim}, N., {et~al.}
  2016{\natexlab{e}}, \aap, 594, A17, \eprint{1502.01592}

\bibitem[{{Planck Collaboration} {et~al.}(2011){Planck Collaboration}, {Ade},
  {Aghanim}, {Arnaud}, {Ashdown}, {Aumont}, {Baccigalupi}, {Balbi}, {Banday},
  {Barreiro}, \& et~al.}]{PlanckCollaboration2011}
{Planck Collaboration}, {Ade}, P.~A.~R., {Aghanim}, N., {et~al.} 2011, \aap,
  536, A8, \eprint{1101.2024}

\bibitem[{{Reichardt} {et~al.}(2013){Reichardt}, {Stalder}, {Bleem}, {Montroy},
  {Aird}, {Andersson}, {Armstrong}, {Ashby}, {Bautz}, {Bayliss}, {Bazin},
  {Benson}, {Brodwin}, {Carlstrom}, {Chang}, {Cho}, {Clocchiatti}, {Crawford},
  {Crites}, {de Haan}, {Desai}, {Dobbs}, {Dudley}, {Foley}, {Forman}, {George},
  {Gladders}, {Gonzalez}, {Halverson}, {Harrington}, {High}, {Holder},
  {Holzapfel}, {Hoover}, {Hrubes}, {Jones}, {Joy}, {Keisler}, {Knox}, {Lee},
  {Leitch}, {Liu}, {Lueker}, {Luong-Van}, {Mantz}, {Marrone}, {McDonald},
  {McMahon}, {Mehl}, {Meyer}, {Mocanu}, {Mohr}, {Murray}, {Natoli}, {Padin},
  {Plagge}, {Pryke}, {Rest}, {Ruel}, {Ruhl}, {Saliwanchik}, {Saro}, {Sayre},
  {Schaffer}, {Shaw}, {Shirokoff}, {Song}, {Spieler}, {Staniszewski}, {Stark},
  {Story}, {Stubbs}, {{\v S}uhada}, {van Engelen}, {Vanderlinde}, {Vieira},
  {Vikhlinin}, {Williamson}, {Zahn}, \& {Zenteno}}]{Reichardt2013}
{Reichardt}, C.~L., {Stalder}, B., {Bleem}, L.~E., {et~al.} 2013, \apj, 763,
  127, \eprint{1203.5775}

\bibitem[{{Rumsey} {et~al.}(2016){Rumsey}, {Olamaie}, {Perrott}, {Russell},
  {Feroz}, {Grainge}, {Handley}, {Hobson}, {Saunders}, \&
  {Schammel}}]{Rumsey2016}
{Rumsey}, C., {Olamaie}, M., {Perrott}, Y.~C., {et~al.} 2016, \mnras, 460, 569,
  \eprint{1604.06120}

\bibitem[{{Sunyaev} \& {Zeldovich}(1970)}]{Sunyaev1970}
{Sunyaev}, R.~A. \& {Zeldovich}, Y.~B. 1970, \apss, 7, 3

\bibitem[{{Vikhlinin} {et~al.}(2006){Vikhlinin}, {Kravtsov}, {Forman}, {Jones},
  {Markevitch}, {Murray}, \& {Van Speybroeck}}]{Vikhlinin2006}
{Vikhlinin}, A., {Kravtsov}, A., {Forman}, W., {et~al.} 2006, \apj, 640, 691,
  \eprint{astro-ph/0507092}

\bibitem[{{Zwart} {et~al.}(2008){Zwart}, {Barker}, {Biddulph}, {Bly}, {Boysen},
  {Brown}, {Clementson}, {Crofts}, {Culverhouse}, {Czeres}, {Dace}, {Davies},
  {D'Alessandro}, {Doherty}, {Duggan}, {Ely}, {Felvus}, {Feroz}, {Flynn},
  {Franzen}, {Geisb{\"u}sch}, {G{\'e}nova-Santos}, {Grainge}, {Grainger},
  {Hammett}, {Hills}, {Hobson}, {Holler}, {Hurley-Walker}, {Jilley}, {Jones},
  {Kaneko}, {Kneissl}, {Lancaster}, {Lasenby}, {Marshall}, {Newton}, {Norris},
  {Northrop}, {Odell}, {Petencin}, {Pober}, {Pooley}, {Pospieszalski}, {Quy},
  {Rodr{\'{\i}}guez-Gonz{\'a}lvez}, {Saunders}, {Scaife}, {Schofield}, {Scott},
  {Shaw}, {Shimwell}, {Smith}, {Taylor}, {Titterington}, {Veli{\'c}},
  {Waldram}, {West}, {Wood}, {Yassin}, \& {AMI Consortium}}]{Zwart2008}
{Zwart}, J.~T.~L., {Barker}, R.~W., {Biddulph}, P., {et~al.} 2008, \mnras, 391,
  1545, \eprint{0807.2469}

\end{thebibliography}
\bibliographystyle{aaeprint}

\end{document}